\documentclass{article}

\usepackage[a4paper]{geometry}
\usepackage{graphicx}
\usepackage{amsmath,amssymb,amsthm}
\usepackage{setspace}
\RequirePackage[hidelinks]{hyperref}
\usepackage{xcolor,colortbl}

\usepackage{tikz}
\usetikzlibrary{shapes.geometric}

\usepackage{natbib}

\newtheorem{theorem}{Theorem}
\newtheorem{corollary}{Corollary}
\newtheorem{lemma}{Lemma}
\newtheorem{assumption}{Assumption}
\newtheorem{definition}{Definition}
\newtheorem{remark}{Remark}

\makeatletter
\newcommand{\tj}[1]{\tag*{\footnotesize #1}}
\makeatother

\newcommand{\tr}{\operatorname{trace}}

\newcommand{\diag}{\operatorname{diag}}
\newcommand{\diagF}[1]{\ensuremath{\diag[#1]}}

\DeclareMathOperator*{\argmin}{arg\,min}

\newcommand{\hr}{H\"usler--Reiss\ }
\newcommand{\hre}{H\"usler--Reiss}

\newcommand{\mones}{{\boldsymbol{1}}}

\renewcommand{\det}{\operatorname{det}}

\newcommand{\norm}[1]{|\!|#1|\!|}
\newcommand{\normB}[1]{\big|\mkern-3.5mu\big|#1\bigr|\mkern-3.5mu\bigr|}

\newcommand{\normf}[1]{|\!|\!|#1|\!|\!|_{\operatorname{F}}}
\newcommand{\normfs}[1]{|\!|\!|#1|\!|\!|^2_{\operatorname{F}}}

\newcommand{\normtwo}[1]{\norm{#1}_2}
\newcommand{\normtwoB}[1]{\normB{#1}_2}
\newcommand{\normtwos}[1]{\norm{#1}_2^2}
\newcommand{\normtwosB}[1]{\normB{#1}_2^2}

\newcommand{\normoneB}[1]{\normB{#1}_1}
\newcommand{\normsup}[1]{\norm{#1}_\infty}

\newcommand{\inprod}[2]{\langle#1,#2\rangle}
\newcommand{\inprodB}[2]{\bigl\langle#1,#2\bigr\rangle}

\newcommand{\inprodBBB}[2]{\biggl\langle#1,#2\biggr\rangle}

\newcommand{\abs}[1]{|#1|}

\newcommand{\R}{\mathbb{R}}
\newcommand{\E}{\mathbb{E}}

\newcommand{\tp}{^\top}
\newcommand{\inv}{^{-1}}

\makeatletter \newcommand*{\deq}{\mathrel{\rlap{%
			\raisebox{0.3ex}{$\m@th\cdot$}}%
		\raisebox{-0.3ex}{$\m@th\cdot$}}=} \makeatother

\newcommand{\dimfull}{{\ensuremath{d}}}
\newcommand{\dimsmall}{{\ensuremath{d-1}}}
\newcommand{\dimsmallb}{{(\ensuremath{d-1})}}

\newcommand{\datafunctionE}{{\ensuremath{\mathfrak{f}_1}}}
\newcommand{\datafunction}{{\ensuremath{\boldsymbol{\mathfrak{f}}_1}}}
\newcommand{\datafunctionA}{{\ensuremath{\datafunction[\argument]}}}
\newcommand{\datafunctionAi}{{\ensuremath{\datafunction[\argumenti]}}}

\newcommand{\datafunctionPE}{{\ensuremath{\mathfrak{f}_2}}}
\newcommand{\datafunctionP}{{\ensuremath{\boldsymbol{\mathfrak{f}}_2}}}
\newcommand{\datafunctionPA}{{\ensuremath{\datafunctionP[\argument]}}}

\newcommand{\datamatrix}{{\ensuremath{\boldsymbol{\boldsymbol{F}}}}}
\newcommand{\datamatrixA}{{\ensuremath{\datamatrix[\argument]}}}

\newcommand{\restrictedeigenvalue}{{\ensuremath{c_{\argument}}}}
\newcommand{\restrictedeigenvalues}{{\ensuremath{c_{\argument}^2}}}

\newcommand{\sparsityV}{{\ensuremath{\mathcal{S}_{\complementaryparameter}}}}
\newcommand{\sparsityM}{{\ensuremath{\mathcal{S}_{\modifiedvariogram}}}}

\newcommand{\sparsityVU}{{\ensuremath{_{\mathcal{S}_{\complementaryparameter}}}}}
\newcommand{\sparsityMU}{{\ensuremath{_{\mathcal{S}_{\modifiedvariogram}}}}}

\newcommand{\sparsityVUC}{{\ensuremath{_{\mathcal{S}_{\complementaryparameter}^\complement}}}}
\newcommand{\sparsityMUC}{{\ensuremath{_{\mathcal{S}_{\modifiedvariogram}^\complement}}}}

\usepackage[bb=dsserif]{mathalpha}
\newcommand{\indicator}[1]{\ensuremath{\mathbb{1}\!\{#1\}}}

\newcommand{\rvY}{{\boldsymbol{Y}}}

\newcommand{\probabilityFB}[1]{\ensuremath{{\mathbb{P}}\bigl\{#1\bigr\}}}

\newcommand{\densitytrad}{{\ensuremath{\mathfrak{g}}}}
\newcommand{\densitytradF}[1]{{\ensuremath{\densitytrad[#1]}}}

\newcommand{\density}{{\ensuremath{\mathfrak{h}}}}
\newcommand{\densityF}[1]{{\ensuremath{\density[#1]}}}

\newcommand{\prior}{{\ensuremath{\mathfrak{p}}}}
\newcommand{\priorF}[1]{{\ensuremath{\prior[#1]}}}

\newcommand{\priorD}{{\ensuremath{\widetilde{\mathfrak{p}}}}}

\newcommand{\priorDFBBB}[1]{{\ensuremath{\widetilde{\prior}\biggl[#1\biggr]}}}
\newcommand{\priorDFBBBB}[1]{{\ensuremath{\widetilde{\prior}\Biggl[#1\biggr]}}}

\newcommand{\constprior}{{\ensuremath{c_{\prior}}}}
\newcommand{\constpriors}{{\ensuremath{c_{\prior}^2}}}

\newcommand{\argumenthigh}{{\ensuremath{\argumentE_{\indexhigh}}}}
\newcommand{\argumenthighM}{{\ensuremath{\argument_{-\indexhigh}}}}
\newcommand{\argumenti}{{\ensuremath{\argument_{i}}}}

\newcommand{\weightE}{{\ensuremath{\mathfrak{w}}}}
\newcommand{\weightEF}[1]{{\ensuremath{\weightE[#1]}}}

\newcommand{\weight}{{\ensuremath{\boldsymbol{\weightE}}}}

\newcommand{\weightA}{{\ensuremath{\weight[\argument]}}}
\newcommand{\weightAE}{{\ensuremath{\weightE[\argument]}}}

\newcommand{\partitiontrad}{{\ensuremath{c_{\variogram}}}}
\newcommand{\partition}{{\ensuremath{c_{\boldsymbol{\mu},\modifiedvariogramL}}}}

\newcommand{\partitionM}{{\ensuremath{c_{\boldsymbol{\mu},\modifiedvariogram}}}}

\newcommand{\covarianceE}{{\ensuremath{\Sigma}}}
\newcommand{\covariance}{{\ensuremath{\boldsymbol{\Sigma}}}}

\newcommand{\numberofobservations}{{n}}
\newcommand{\tuningparameter}{{r}}
\newcommand{\tuningparameterO}{{\tuningparameter^*}}
\newcommand{\tuningparameterOA}{{\tuningparameterO[\complementaryparameter,\modifiedvariogram]}}
\newcommand{\scorefunction}{{\boldsymbol{\mathfrak{s}}}}
\newcommand{\scorefunctionE}{{\mathfrak{s}}}
\newcommand{\objectivefunction}{{\mathfrak{j}}}
\newcommand{\objectivefunctionI}{{\mathfrak{o}}}

\newcommand{\empiricaldensity}{{\ensuremath{\mathfrak{p}}}}

\newcommand{\empiricaldensityF}[1]{{\ensuremath{\empiricaldensity[#1]}}}

\newcommand{\modifiedvariogramE}{{\Theta}}
\newcommand{\modifiedvariogram}{{\boldsymbol{\Theta}}}
\newcommand{\modifiedvariogramLE}{{\Lambda}}
\newcommand{\varioextE}{{\diagF{\modifiedvariogramL\mones+\modifiedvariogramL\tp\mones}}}
\newcommand{\varioext}{{-\diagF{\modifiedvariogramL\mones+\modifiedvariogramL\tp\mones}}}
\newcommand{\varioextest}{{\diagF{\modifiedvariogramLest\mones+\modifiedvariogramLest\tp\mones}}}
\newcommand{\modifiedvariogramL}{{\boldsymbol{\Lambda}}}
\newcommand{\complementaryparameterE}{{\ensuremath{\mu}}}
\newcommand{\complementaryparameter}{{\ensuremath{\boldsymbol{\mu}}}}
\newcommand{\complementaryparameterest}{{\ensuremath{\widehat{\complementaryparameter}}}}

\newcommand{\modifiedvariogramLest}{{\widehat{\modifiedvariogramL}}}
\newcommand{\modifiedvariogramest}{{\widehat{\modifiedvariogram}}}

\newcommand{\locationS}{{\ensuremath{\mathcal{L}}}}
\newcommand{\variogrammatrices}{{\ensuremath{\mathcal{W}}}}
\newcommand{\variogrammatricesS}{{\ensuremath{\mathcal{V}}}}

\newcommand{\nbrdim}{\ensuremath{d}}
\newcommand{\nbrdimU}{\ensuremath{_{\nbrdim}}}
\newcommand{\nbrdimR}{\ensuremath{\nbrdim-1}}
\newcommand{\nbrdimRQ}{\ensuremath{(\nbrdimR)}}
\newcommand{\nbrsamples}{\ensuremath{n}}

\newcommand{\Rdim}{\ensuremath{\R^{\nbrdim}}}
\newcommand{\RdimR}{\ensuremath{\R^{\nbrdimR}}}
\newcommand{\Rdimdim}{\ensuremath{\R^{\nbrdim\times\nbrdim}}}
\newcommand{\RdimRdimR}{\ensuremath{\R^{(\nbrdimR)\times(\nbrdimR)}}}

\newcommand{\variogram}{\ensuremath{\boldsymbol{\Gamma}}}
\newcommand{\variogramE}{\ensuremath{\Gamma}}
\newcommand{\variogramS}{\ensuremath{\mathcal{G}}}
\newcommand{\variogramest}{\ensuremath{\boldsymbol{\widehat\variogram}}}
\newcommand{\variogramestE}{\ensuremath{\widehat\variogramE}}

\newcommand{\indexhigh}{\ensuremath{m}}

\newcommand{\indexsample}{\ensuremath{\ensuremath{i}}}

\newcommand{\indexelement}{\ensuremath{j}}
\newcommand{\indexelementP}{\ensuremath{k}}
\newcommand{\indexelementPP}{\ensuremath{l}}

\newcommand{\normaldensity}{\ensuremath{\mathfrak{n}_{\nbrdimR}}}

\newcommand{\normaldensityFB}[1]{\ensuremath{\normaldensity\bigl[#1\bigr]}}

\newcommand{\argumentE}{\ensuremath{x}}
\newcommand{\argument}{\ensuremath{\boldsymbol{\argumentE}}}

\newcommand{\maxargumentS}{\ensuremath{\mathcal{S}}}

\newcommand{\rv}{\ensuremath{\ensuremath{\boldsymbol{X}}}}
\newcommand{\rvyE}{\ensuremath{\ensuremath{Y}}}

\newcommand{\rvZE}{\ensuremath{\ensuremath{Z}}} 
\newcommand{\rvZ}{\ensuremath{\ensuremath{\boldsymbol{\rvZE}}}}

\newcommand{\rvWE}{\ensuremath{\ensuremath{W}}} 
\newcommand{\rvW}{\ensuremath{\ensuremath{\boldsymbol{\rvWE}}}}

\usepackage{authblk}

	\title{Extremes in High Dimensions: Methods and Scalable Algorithms}

\author[1]{Johannes Lederer}
\affil[1]{University of Hamburg, \texttt{johannes.lederer@uni-hamburg.de}}
\author[2]{Marco Oesting}
\affil[2]{University of Stuttgart, \texttt{marco.oesting@mathematik.uni-stuttgart.de}}

\date{}

\begin{document}

	\maketitle

	\begin{abstract}
		Extreme value theory for univariate and low-dimensional observations has been explored in considerable detail, but the field is still in an early stage regarding high-dimensional settings. This paper focuses on H\"usler-Reiss models, a popular class of models for multivariate extremes similar to multivariate Gaussian distributions, and their domain of attraction. We develop estimators for the model parameters based on score matching, and we equip these estimators with theories and exceptionally scalable algorithms. Simulations and applications to weather extremes demonstrate the fact that the estimators can estimate a large number of parameters reliably and fast; for example, we show that H\"usler-Reiss models with thousands of parameters can be fitted within a couple of minutes on a standard laptop. More generally speaking, our work relates extreme value theory to modern concepts of high-dimensional statistics and convex optimization.
	\end{abstract}

\section{Introduction}

Low-dimensional models for extremes and corresponding methods and theories have been studied extensively~\citep{Embrechts2013,Haan2006,Resnick2008}.
Complex extremal events, such as global weather phenomena, however, require much more detailed and flexible models, particularly high-dimensional models with many estimable parameters.

In general, high-dimensional models are abundant in statistics and machine learning~\citep{Lederer2022,Buhlmann2011}. 
For extremes, however, high-dimensional data are currently analyzed mainly in the context of spatial or spatio-temporal extremes \citep[see][for recent reviews]{engelke-ivanovs-21,huser-wadsworth-22} with models such as the \hr model \citep{Husler1989}.
This model, and its spatial and spatio-temporal extensions, the Brown--Resnick processes \citep{KSH-09}, are frequently used in a wide range of applications \citep[see][for instance]{gaume-etal-13, ADE-15, buhl-klueppelberg-16, thibaud-etal-16, OSF-17, oesting-stein-18, EFO-19}. In most of these applications, similarly to typical approaches in classical geostatistics, the underlying variogram of the model is parametrized as a (in most cases relatively simple) function of spatial coordinates or time and, consequently, fitting such a model boils down to the estimation of a small number of parameters.

Using an adapted definition of conditional independence for extremes \citep{Engelke2020}, 
the \hr models can be shown to exhibit many similarities with Gaussian graphical models, which make them particularly promising in high-dimensional settings with a large number of estimable parameters. So far, several interesting approaches to high-dimensional \hr models exist, but they have severe limitations.
Approaches based on total positivity~\citep{2021Roettger} are very restrictive.
The EGlearn algorithm~\citep{Engelke2022} does away with this restriction.
Given $\nbrdim$-dimensional samples, its course of action is as follows: 
estimate the variogram, then apply a ``base learner'' to $\nbrdim$~variants of this estimate,
and eventually combine the $\nbrdim$~outputs of the base learner via majority vote.
But this approach has a number of other limitations:
(i)~It only estimates the edges of a graph and, therefore, does not yield a generative model.
(ii)~It is unclear what base learner is appropriate; 
for example, base learners developed for Gaussian data may be consistent but not necessarily efficient.
(iii)~The fact that the base learner needs to be applied~$\nbrdim$ times renders the approach infeasible for large~$\nbrdim$;
and even if the approach is feasible,
the repeated application of the base learner inflicts theoretical challenges (for example, $\nbrdim$ sets of assumptions are needed) and practical challenges (for example, $\nbrdim$ tuning parameters are needed when the graphical lasso is used as the base learner and even $\nbrdim(\nbrdim-1)$ tuning parameters when neighborhood selection is used).
In brief, the EGlearn algorithm is an interesting step forward,
but it still suffers from a number of issues.

This paper approaches data from the \hr model or its domain of attraction very differently.
We first disentangle the model parameters, 
which resemble the mean and the precision matrix of a log-Gaussian density, by generalizing the \hr densities.
We then introduce parameter estimators that are motivated by score matching~\citep{Hyvarinen2005} and recent extensions of it~\citep{Liu2019,Yu2020}.
The main feature of score matching is that it involves the derivatives of log-densities rather than the densities themselves,
which circumvents the time-consuming, if not impossible, calculation of normalization constants.
Using this, we can formulate our estimators as minimizers of objective functions that are convex and benign more generally.
Thus, while known approaches to multivariate extremes are not scalable
(in fact, accurate computations might not even be ensured in low dimensions),
our estimators are amenable to highly efficient algorithms from convex optimization.

The comparably simple form of our estimators also facilitates the development of statistical theory and regularization by including prior terms that induce sparsity or other structures in the parameter space. In other words, our framework can also leverage the strengths of high-dimensional statistics.
	
\paragraph*{Outline}

The remainder of this section discusses other related papers and sets some notation.
Section~\ref{sec:method} introduces the framework and the estimators. Section~\ref{sec:theory} establishes statistical theory, including guarantees for high-dimensional settings,
where the number of parameters~$\nbrdim+\nbrdim\nbrdimRQ/2$ is large, potentially much larger than the number of samples~$\nbrsamples$.
Section~\ref{sec:numerical} demonstrates the performance of our approach in practice including both simulated and real data.
Section~\ref{sec:discussion} provides a short discussion.
Further technical results and all proofs are deferred to the supplementary materials.
	
\paragraph*{Further related literature}

High-dimensional data, particularly spatially or spatio-tem\-porally indexed data, have become an increasingly popular topic in the field of extreme value statistics.
Parametric modeling of their dependence structure is so far often accompanied by additional assumptions such as stationarity in time, isotropy in space, space-time separability, or non-stationarity structures that can be traced back to a small number of covariates \citep[see][for instance]{buhl-etal-19, hazra-huser-21, hazra-etal-21, simpson-wadsworth-21, forster-oesting-22}. Such approaches based on a certain form of homogeneity over the spatio-temporal domain are complemented by concepts based on localized likelihood structures 
\citep{castrocamilo-huser-20,shao-etal-22} or on a recently introduced form of conditional independence tailored to extremes \citep{Engelke2020}. Exploiting such structures for statistical inference, for example, by fitting a (composite/censored) likelihood \citep[see][for instance]{padoan-etal-10, huser-etal-19} or by gradient scoring \citep{Fondeville2018, defondeville-davison-22}, results in low- or moderate-dimensional optimization problems.
This allows for the application of standard optimization techniques in principle,
but the typical non-convexity of those optimization problems can make the results  highly sensitive to starting values and the specification of the optimization procedure---see the discussion in \cite{forster-oesting-22}, for instance.
	
\paragraph*{Notation}

The vector $\argument_{-\indexhigh}\in\RdimR$ is $\argument = (\argument_{1},\ldots, \argument_{\dimfull})^\top \in\Rdim$ with the $\indexhigh$th element removed.
Similarly, the vector $\boldsymbol{A}_{-\indexhigh,\indexhigh}\in\RdimR$ is the $\indexhigh$th column of $\boldsymbol{A}\in\Rdimdim$ with the $\indexhigh$th element removed.
The matrix $\diagF{\argument}\in\Rdimdim$ for given $\argument\in\Rdim$ is defined through $(\diagF{\argument})_{\indexelement,\indexelementP}\deq\argumentE_{\indexelement}$ if $\indexelement=\indexelementP$ and $(\diagF{\argument})_{\indexelement,\indexelementP}\deq0$ otherwise. The Frobenius norm of a matrix~$\boldsymbol{A}\in\Rdimdim$ is $\normf{\boldsymbol{A}}\deq\sqrt{\sum_{\indexelement,\indexelementP=1}^{\nbrdim}(A_{\indexelement\indexelementP})^2}$. All integrals are Lebesgue integrals.
The Lebesgue density of the normal distribution in~$\RdimR$ with mean vector~$\boldsymbol{\mu}\in\RdimR$ and covariance matrix~$\boldsymbol{\Sigma}\in\RdimRdimR$ is denoted by $\normaldensity[\,\,\cdot\,\,;\boldsymbol{\mu},\boldsymbol{\Sigma}]$.
Expressions such as $\rv>\boldsymbol{x}$ for (same length) vectors $\rv,\boldsymbol{x}$ are understood component-wise.

\section{Method}
	\label{sec:method}
	
\subsection{Multivariate Regular Variation}
	\label{mrv}

A common assumption in multivariate extreme value statistics is the multivariate regular variation of the random vector $\rv \in \Rdim$ of interest. After some marginal transformations, without loss of generality, we may assume that $\rv > 0$ and that $\probabilityFB{X_i > x} \sim c_i/x$ as $x \to \infty$ for appropriate constants $c_i>0$ and $i\in\{1,\ldots,\nbrdim\}$. Then, the vector $\rv$ is called regularly varying if
$$ \lim_{t \to \infty} \frac{1 - \probabilityFB{\rv \leq t\boldsymbol{x}}}{1 - \probabilityFB{\rv \leq t\mones}} = \mu\bigl([\boldsymbol{0},\boldsymbol{x}]^\complement\bigr) $$
for all $\boldsymbol{x} \in[0,\infty)^{\nbrdim} \setminus \{\boldsymbol{0}\}$ and some measure $\mu$ on $(0,\infty)^{\nbrdim} \setminus \{\boldsymbol{0}\}$ that is homogeneous of order~$-1$, that is,
$ t \mu(t\mathcal{A}) = \mu(\mathcal{A})$ for every $t>0$ and every measurable subset $ \mathcal{A} \subset[0,\infty)^{\nbrdim} \setminus \{\boldsymbol{0}\}$ bounded away from $\boldsymbol{0}$ (see, e.g., \citealp{Resnick2008}). The measure $\mu$ is called the exponent measure.
Multivariate regular variation is equivalent to
\begin{equation} \label{eq:lim-maxstable}
	\lim_{n \to \infty} \probabilityFB{\max\{\rv_1, \ldots, \rv_n\}/n \leq \boldsymbol{x}} = \exp\bigl[-\mu([\boldsymbol{0},\boldsymbol{x}]^\complement)\bigr] 
\end{equation}
for all $\boldsymbol{x} \in[0,\infty)^{\nbrdim} \setminus \{\boldsymbol{0}\}$ where $\rv_1,\ldots,\rv_n$ are independent copies of the vector $\rv$ and the maximum is taken componentwise. Note that the limit in \eqref{eq:lim-maxstable} defines a c.d.f.~on $[0,\infty)^{\nbrdim} \setminus \{\boldsymbol{0}\}$, a so-called multivariate max-stable distribution. The convergence in \eqref{eq:lim-maxstable} forms the basis for many so-called block maxima approaches in extreme value statistics.

Besides block maxima, inference for multivariate extremes is often based on peaks-over-threshold approaches. These make use of the fact that multivariate regular variation implies that,
for any $u\in(0,\infty)$ and norm~$\norm{\!\;\cdot\!\;}$ on~$\R^{\dimfull}$,
\begin{equation} \label{eq:lim-pot}
	\lim_{u \to \infty} \probabilityFB{\rv /u \in \mathcal{A} \mid \norm{\rv} > u} = \frac{\mu(\mathcal{A})}{\mu(\maxargumentS)} 
\end{equation}
for every measurable subset $\mathcal{A} \subset \maxargumentS \deq \{ \argument \in[0,\infty)^{\dimfull}\,: \, \norm{\argument} > 1 \}$ such that $\mu(\partial \mathcal{A})=0$. The right-hand side of \eqref{eq:lim-pot} defines a so-called multivariate Pareto distribution.

In view of Equations \eqref{eq:lim-maxstable} and \eqref{eq:lim-pot}, 
parametric models for multivariate max-stable and multivariate Pareto distributions are usually based on parametric models for the exponent measure~$\mu$. One of the most popular parametric models is the \hr model~\citep{Husler1989} discussed in the following subsection.
	
\subsection{The \hr Model}
\label{HueslerReiss}

We consider symmetric, conditionally strictly negative-definite matrices $\variogram\in\Rdimdim$, i.e.,
\begin{align*}
	\variogram~\in~\variogramS~\deq{}&~\{\boldsymbol{A}\in\Rdimdim\,:\,\boldsymbol{A}\tp=\boldsymbol{A}, \diagF{\boldsymbol{A}}=\boldsymbol{0}~\text{and}\\
	&~\boldsymbol{c}\tp \boldsymbol{A}\boldsymbol{c}< 0~\text{for all}~\boldsymbol{c}\in\Rdim\setminus\{\boldsymbol{0}\nbrdimU\}~\text{that satisfy}~c_1+\dots+c\nbrdimU=0\}\,,
\end{align*}
compare to \citet[p.~66]{Berg-etal-1984}, for example. Such matrices are valid variogram matrices, that is, for each $\variogram \in \variogramS$, there exists a centered Gaussian random vector $\rvW \in \Rdim$ such that 
$\variogram = \left( \mathrm{Var}(\rvWE_i -\rvWE_j) \right)_{1 \leq i,j \leq \nbrdim}$.
Based on the distribution of $\rvW$, we can define a valid exponent measure (up to some normalizing constant that can be chosen according to the desired standardization) via
\begin{equation} \label{eq:exp-measure-stochrepr}
	\mu_{\variogram}(\mathcal{A})~\propto~ \int_0^\infty y^{-2} \mathbb{P}\bigl\{y\exp(\rvW - \mathrm{Var}(\rvW)/2) \in \mathcal{A}\bigr\}  \,\mathrm{d}y, \quad
	\mathcal{A} \subset [0,\infty)^{\nbrdim} \setminus \{\boldsymbol{0}\}, 
\end{equation}
where $\mathrm{Var}(\rvW) = (\mathrm{Var}(\rvWE_i))_{1 \leq i \leq \nbrdim}$ denotes the vector of componentwise variances. Extreme value models based on this exponent measure
are called \hr models associated with the variogram matrix $\variogram$. More precisely, the corresponding max-stable distribution defined by the limit in \eqref{eq:lim-maxstable} is called max-stable \hr distribution, while the corresponding multivariate Pareto model defined by the right-hand side of \eqref{eq:lim-pot} is called \hr Pareto model. 
Hence, a random vector~$\rv\in\Rdim$ is said to follow the \hr Pareto model if the equality in~\eqref{eq:lim-pot} already holds for every finite~$u$. 

The representation of the \hr exponent measure $\mu_{\variogram}$ in \eqref{eq:exp-measure-stochrepr} turns out to be of limited use for statistical inference, 
but a more convenient representation can be obtained by the following observation: While the law of $\rvW$ is not uniquely defined by $\variogram$, the exponent measure $\mu_{\variogram}$ is. In other words: We can choose any centered Gaussian random vector~$\rvW$ with variogram
matrix $\variogram$ to calculate $\mu_{\variogram}$ in \eqref{eq:exp-measure-stochrepr}.
In particular, we can fix an arbitrary index~$\indexhigh\in\{1,\dots,\dimfull\}$ and
consider the uniquely defined centered Gaussian vector $\rvW$ with variogram matrix $\variogram$ and $\rvWE_{\indexhigh}=0$ almost surely. By a straightforward calculation, one can see that the $(\nbrdim-1)$-dimensional Gaussian vector $\rvW_{-\indexhigh}$ has the covariance matrix~$\covariance\equiv\covariance[\indexhigh]\in\R^{\dimsmallb\times\dimsmallb}$ defined through
\begin{equation} \label{eq:vario2covar}
	\covarianceE_{kl}~\deq~
	\begin{cases}
		(\variogramE_{k\indexhigh} + \variogramE_{\indexhigh l} - \variogramE_{kl})/2&\text{if}~k,l<\indexhigh\\
		(\variogramE_{k\indexhigh} + \variogramE_{\indexhigh(l+1)} - \variogramE_{k(l+1)})/2&\text{if}~k<\indexhigh,l\geq\indexhigh\\
		(\variogramE_{(k+1)\indexhigh} + \variogramE_{\indexhigh l} - \variogramE_{(k+1)l})/2&\text{if}~k\geq\indexhigh,l<\indexhigh\\
		(\variogramE_{(k+1)\indexhigh} + \variogramE_{\indexhigh(l+1)} - \variogramE_{(k+1)(l+1)})/2&\text{if}~k,l\geq\indexhigh
	\end{cases}
\end{equation}
for all~$k,l\in\{1,\dots,\dimsmall\}$. Thus, \covariance\ is a covariance matrix corresponding to the variogram matrix \variogram---see, for example, \cite{pistone2020model}. It can be shown that~\covariance\ is positive definite, that is, the vector $\rvW_{-\indexhigh}$ possesses a $(\nbrdim-1)$-dimensional Lebesgue density. Using this density, it can be shown that the exponent measure $\mu_{\variogram}$ in \eqref{eq:exp-measure-stochrepr} is absolutely continuous with respect to the $\nbrdim$-dimensional Lebesgue measure. More precisely, according to \cite{Engelke2015},
$$ \mu_{\variogram}(\mathcal{A})~=~\int\nolimits_{\mathcal{A}}\densitytradF{\argument;\variogram}\,d\argument, \quad
\mathcal{A} \subset [0,\infty)^{\nbrdim} \setminus \{\boldsymbol{0}\}, $$
where $\densitytrad\,:\, [0,\infty)^{\nbrdim} \setminus \{\boldsymbol{0}\}\to(0,\infty]$ is a Lebesque density on the set $[0,\infty)^{\nbrdim} \setminus \{\boldsymbol{0}\}$ given by
\begin{equation} \label{eq:spectral-density}
	\densitytradF{\argument;\variogram}~\deq~\frac{1}{\partitiontrad} \frac{1}{\argumenthigh}\left(\prod\nolimits_{\indexelement=1}^{\dimfull}\frac{1}{\argumentE_{\indexelement}}\right)\normaldensityFB{\log[\argumenthighM/\argumenthigh];-\variogram_{-\indexhigh,\indexhigh}/2,\covariance}
\end{equation}
for all~$\argument\in\maxargumentS$,\label{maxset}
where~$\partitiontrad$ is the normalizing constant
\begin{equation*}
	\partitiontrad~\deq~\int_{\maxargumentS}  \frac{1}{\argumenthigh}\left(\prod\nolimits_{\indexelement=1}^{\dimfull}\frac{1}{\argumentE_{\indexelement}}\right)\normaldensityFB{\log[\argumenthighM/\argumenthigh];-\variogram_{-\indexhigh,\indexhigh}/2,\covariance}\,\mathrm{d}\argument
\end{equation*}
such that the restriction of $\densitytradF{\cdot;\variogram}$ to $\maxargumentS$ defines a probability density.

This representation allows one to write the \hr Pareto distribution as defined above in a convenient way. 
More precisely, $\rv$ follows a \hr Pareto distribution associated with the variogram matrix \variogram\ if $\rv$ satisfies
\begin{equation} \label{eq:eq-pot}
	\probabilityFB{\rv/u \in \mathcal{A} \mid \norm{\rv} > u }~=~\int_{\mathcal{A}}\densitytradF{\argument;\variogram}\,\mathrm{d}\argument
\end{equation}
for all~$u\in(0,\infty)$ and measurable~$\mathcal{A} \subset \maxargumentS$
such that $\mu(\partial \mathcal{A})=0$ \citep{Engelke2015}.

\hr models are widespread in many applications such as financial science and environmental statistics, like statistical investigations of rainfall, temperature, wind speed, and river discharge, where it is natural to assume that the data is normally distributed~\citep{aloui2011global, ADE-15, buhl-klueppelberg-16, davison2013geostatistics, joe1994multivariate}. In these cases, it can be argued the limits of appropriately rescaled maxima of observations follow a max-stable \hr distribution~\citep{Husler1989}. This motivates the use of max-stable \hr and, equivalently for exceedances, \hr Pareto distributions. In the following, we will focus on peaks-over-threshold approaches and, consequently, consider \hr models restricted to $\maxargumentS$. A typical choice for the norm in the definition of~\maxargumentS\ is $\norm{\!\;\cdot\!\;}=\normsup{\!\;\cdot\!\;}$.
We will exemplify our approach for this choice later in the numerical section (Section \ref{sec:numerical}), but we keep our method and theories completely general otherwise.

\subsection{Approximating the \hr Model}
	\label{generalizedmodel}

A challenge of the \hr model is that the means and covariances of the underlying normal distribution are entangled in an intricate way.
In the following,
we propose disentangling these two aspects by generalizing the densities.
In this process, 
we will rewrite the density $\densitytradF{\argument;\variogram}$ in 
\eqref{eq:spectral-density} in a way that is independent of the choice of the index $\indexhigh$, replacing the precision matrix $\covariance^{-1}$ by the \hr precision matrix $\modifiedvariogram$ introduced in \cite{HES-22}.
More precisely, we consider a matrix $\modifiedvariogram$ defined by
\begin{equation} \label{eq:def-modifiedvariogram}
	\modifiedvariogramE_{\indexelement\indexelementP}~\deq~
	\begin{cases}
		(\covarianceE\inv)_{\indexelement\indexelementP} ~&\text{for}~\indexelement,\indexelementP<\indexhigh\,;\\
		(\covarianceE\inv)_{\indexelement(\indexelementP-1)} ~&\text{for}~\indexelement<\indexhigh<\indexelementP\,;\\
		(\covarianceE\inv)_{(\indexelement-1)\indexelementP} ~&\text{for}~\indexelementP<\indexhigh<\indexelement\,;\\
		-\sum_{\indexelementPP=1}^{\dimfull-1}(\covarianceE\inv)_{\indexelementPP\indexelementP}~&\text{for}~\indexelement=\indexhigh, \indexelementP<\indexhigh\,;\\
		-\sum_{\indexelementPP=1}^{\dimfull-1}(\covarianceE\inv)_{\indexelementPP(\indexelementP-1)}~&\text{for}~\indexelement=\indexhigh, \indexelementP>\indexhigh\,;\\
		-\sum_{\indexelementPP=1}^{\dimfull-1}(\covarianceE\inv)_{\indexelement\indexelementPP}~&\text{for}~\indexelement<\indexhigh,\indexelementP=\indexhigh\,;\\
		-\sum_{\indexelementPP=1}^{\dimfull-1}(\covarianceE\inv)_{(\indexelement-1)\indexelementPP}~&\text{for}~\indexelement>\indexhigh,\indexelementP=\indexhigh\,;\\
		\sum_{\indexelementPP_1=1}^{\dimfull-1} \sum_{\indexelementPP_2=1}^{\dimfull-1} (\covarianceE\inv)_{\indexelementPP_1\indexelementPP_2}  ~&\text{for}~\indexelement=\indexelementP=\indexhigh\,;\\
		(\covarianceE\inv)_{(\indexelement-1)(\indexelementP-1)} ~&\text{for}~\indexelement,\indexelementP>\indexhigh\,.
	\end{cases}
\end{equation}
In other words: $\modifiedvariogram$ is constructed by setting $\modifiedvariogram_{-\indexhigh,-\indexhigh} = \covariance\inv$ and filling the $\indexhigh$th row and column in a unique way such that all row sums and column sums are equal to zero. By Lemma 1 in \cite{Engelke2020}, the matrix $\modifiedvariogram$ is independent of the specific choice of $\indexhigh \in \{1, \ldots, \dimfull\}$.
By construction, $\modifiedvariogram$ is symmetric and positive semi-definite, has vanishing row sums and has rank $\dimfull - 1$,
that is, it is an element of the set 
\begin{equation*}
	\variogrammatrices^*\deq\{\boldsymbol{A}\in\variogrammatrices\,:\,\mathrm{rank}(\boldsymbol{A})=\dimfull-1, \boldsymbol{A} \text{ is positive semi-definite}\},
\end{equation*}	
where $\variogrammatrices\deq\{\boldsymbol{A}\in\R^{\dimfull\times\dimfull}\,:\,\text{$\boldsymbol{A}$ is symmetric and has zero-valued row and column sums}\}$.

By Proposition 3.4 in \cite{HES-22}, there is a one-to-one correspondence between the set $\variogramS$ of all potential variogram matrices and the set $\variogrammatrices^*$ of all potential \hr precision matrices, that is, every matrix $\modifiedvariogram \in \variogrammatrices$ with $k-1$ positive eigenvalues corresponds to a valid \hr model and vice versa. This observation suggests formulating the \hr density in terms of $\modifiedvariogram$
as in the following lemma, see also Proposition 3.6 in \citealp{HES-22}. 

\begin{lemma}[Reparametrization via $\modifiedvariogram$]
	\label{reparametrization}
	Given a variogram matrix $\variogram\in\variogramS$ and a corresponding covariance matrix $\covariance\equiv\covariance[\indexhigh]$,
	define $\complementaryparameter\in\R^{\dimfull}$ through
	\begin{equation*}
		\complementaryparameterE_{\indexelement}~\deq\,
		\begin{cases}
			-(\covariance\inv\variogram_{-\indexhigh,\indexhigh})_{\indexelement}/2~~~&\text{for}~\indexelement\neq\indexhigh\\
			\sum_{\indexelementP\neq\indexhigh}(\covariance\inv\variogram_{-\indexhigh,\indexhigh})_{\indexelementP}/2-1~~~&\text{for}~\indexelement=\indexhigh
		\end{cases}
	\end{equation*}
	and a corresponding $\modifiedvariogram \in \variogrammatrices^*$.
	Set further 
	\begin{equation*}
		\partitionM~\deq~\int_{\maxargumentS}  \left(\prod\nolimits_{\indexelement=1}^{\dimfull}\frac{1}{\argumentE_{\indexelement}}\right)\exp\biggl[\complementaryparameter\tp\log[\argument]
		-\frac{1}{2}\log[\argument]\tp \modifiedvariogram \log[\argument]\biggr] \,\mathrm{d}\argument\,.
	\end{equation*}
	Then, $\partitionM<\infty$ is well-defined, and it holds
	\begin{equation*}
		\densitytradF{\argument;\variogram} ~=~ \frac 1 {\partitionM} \left(\prod\nolimits_{\indexelement=1}^{\dimfull}\frac{1}{\argumentE_{\indexelement}}\right)\exp\biggl[\complementaryparameter\tp\log[\argument]
		-\frac{1}{2}\log[\argument]\tp \modifiedvariogram \log[\argument]\biggr], \quad \argument\in\maxargumentS.
	\end{equation*}
\end{lemma}

\begin{remark}[Independence of $\indexhigh$]
	As explained above, the exponent measure $\mu_{\variogram}$ does not depend on the specific choice of $\indexhigh \in \{1,\ldots,\dimfull\}$, and, consequently, the same holds true for its Lebesgue density
	$\densitytradF{\,\cdot\,;\variogram}$. Given the expression in Lemma~\ref{reparametrization}, this confirms that the \hr precision matrix $\modifiedvariogram$ is independent of the choice of $\indexhigh$, and so is the vector $\complementaryparameter$ defined in Lemma \ref{reparametrization}.
\end{remark}

We now consider a more general function class. 
To this end, we first need some more notation.
We consider  sets (one can also take nonempty subsets of them)
$\locationS~\deq~\R^{\dimfull}$ and $\variogrammatricesS~\deq~\bigl\{\boldsymbol{A}\in\R^{\dimfull\times\dimfull}\,:\,A_{ij}=0~\text{for all}~i,j\in\{1,\dots,\dimfull\}~\text{such that}~i\geq j\bigr\}$\,.
Thus, $\locationS$ are simply real-valued vectors in $\dimfull$ dimensions,
whereas $\variogrammatricesS$ are the strictly upper-triangular matrices  in $\dimfull\times\dimfull$ dimensions.
For us, these matrices serve as convenient representations of the matrices that are symmetric and have all row and column sums equal to zero:
indeed, every $\modifiedvariogramL\in\variogrammatricesS$ yields a $\modifiedvariogram\in\variogrammatrices$ via\label{thetadef}
\begin{equation*}
	\modifiedvariogram~=~\modifiedvariogramL+\modifiedvariogramL\tp\varioext\,,
\end{equation*}
and vice versa.
We will use~$\modifiedvariogramL$ and~$\modifiedvariogram$ interchangeably in our notation.
In our generalized models, the elements in $\locationS$ and $\variogrammatrices$ (or, equivalently $\variogrammatricesS$) will play similar roles as mean vectors and precision matrices in standard (log-)Gaussian distributions.

Given $\complementaryparameter$ and $\modifiedvariogramL$, 
we then  define a function $\density\,:\,\maxargumentS\to[0,\infty)$ by
\begin{multline}
	\label{newdensity}
	\densityF{\argument;\complementaryparameter,\modifiedvariogramL}~\deq~\frac{1}{\partition}\left(\prod\nolimits_{\indexelement=1}^{\dimfull}\frac{1}{\argumentE_{\indexelement}}\right)\exp\biggl[\complementaryparameter\tp\log[\argument]\\
	-\frac{1}{2}\tr\Bigl[(\modifiedvariogramL+\modifiedvariogramL\tp\varioext)\log[\argument]\log[\argument]\tp\Bigr] \biggr]
\end{multline}
for all~$\argument\in\maxargumentS$ and an arbitray factor $\partition\in(0,\infty)$ that does not depend on~$\argument$.
These functions will be used to approximate the \hr densities---or more precisely, the shape of those densities. The functions \density\ indeed generalize the traditional \hr densities \densitytrad\ as has been shown in Lemma \ref{reparametrization}.
However, we do not need \density\ to be a density in general---see the  following section. 
The  key advantage of the function~\density\ is that the parameters $\complementaryparameter$ and $\modifiedvariogramL$ are disentangled entirely.

Note that the dimension of the parameter space~$\locationS\times\variogrammatricesS$ is $\nbrdim+\nbrdim\nbrdimRQ/2$, that is, the number of parameters in the \hr model and in its generalization increases quadratically in the dimension~\nbrdim\ of the samples.
This relationship highlights the network flavor of the \hr model,
and it emphasizes the fact that high-dimensional techniques are needed even for moderate~\nbrdim.

\subsection{Score-Matching Estimator} \label{subsec:scorematching}

It is tempting to try estimating the parameters in the \hr model by maximum likelihood.
Such an approach seems particularly attractive because there  are many established theories for maximum likelihood in statistics.
But even if extensive theories could be developed in the \hr framework, 
they would not bear practical significance because the likelihood involves the normalization constant~$\partitiontrad$, which is extremely hard to compute.
We, therefore, propose a different approach based on our generalizations of the density and on score matching. Our approach does not involve any normalization constants,
and it is computationally scalable more generally.

As in score-matching, we start with the score functions
\begin{align*}
	\scorefunction[\,\cdot\,;\complementaryparameter,\modifiedvariogramL]~:~\maxargumentS\,&\to\,\mathbb{R}^{\dimfull}, \
	\argument\,\mapsto\,\nabla_{\argument} \log\bigl[\densityF{\argument;\complementaryparameter,\modifiedvariogramL}\bigr]
\end{align*}
for all $\complementaryparameter\in\locationS,\modifiedvariogramL\in\variogrammatricesS$.
But in our case, the score functions are not score functions in a strict sense 
because we do not require the functions $\density$ to be densities.
The idea is that (i)~normalization constants and other factors that only depend on the argument~$\argument$ do not affect the score anyway and that (ii)~understanding the dependence structures of the data does not require the whole density but only the ``shape'' of it.

We then use the score functions to assess the quality of the parameters $\complementaryparameter$ and $\modifiedvariogramL$.
Standard score matching measures the parameters' quality by \citep[p.~697ff]{Hyvarinen2005}
\begin{equation*}
	\frac{1}{2} \int_{\maxargumentS}\empiricaldensityF{\argument}\normtwosB{\scorefunction[\argument;\complementaryparameter, \modifiedvariogramL]-\scorefunction[\argument]} \mathrm{~d} \argument
\end{equation*}
where $\empiricaldensity$ is the data-generating density and $\scorefunction[\argument]\deq \nabla_{\argument} \log \empiricaldensityF{\argument}$  are the values of the ``true'' score function.
But this is not possible here
because this function does not yield the desired reformulation that we derive later in Theorem~\ref{scorematchinggeneral}.
Inspired by follow-up work to \citet{Hyvarinen2005},
such as \citet{Hyvarinen2007,Liu2019,Yu2020},
we generalize the score-matching approach to
\begin{equation}\label{scalingscorematching}
	\objectivefunction [\complementaryparameter,\modifiedvariogramL]~\deq~\frac{1}{2} \int_{\maxargumentS} \empiricaldensityF{\argument}\normtwoB{\scorefunction [\argument] \otimes \argument\otimes\weightA-\scorefunction[\argument ; \complementaryparameter,\modifiedvariogramL] \otimes \argument \otimes\weightA}^{2} \mathrm{~d} \argument\,,
\end{equation}
where
$\otimes$ denotes element-wise multiplication,
	$\weightA = (\weightE[\argument_1],\ldots, \weightE[\argument_{\dimfull}])^\top$ for some differentiable weight function $\weightE\,:\,\R\to\R$,
and $\complementaryparameter\in\locationS,\modifiedvariogramL\in\variogrammatricesS$.

Note that the ``true,'' data-generating score function does not need to correspond to a \hr model,
that is, we do not assume that the data follow a \hr model exactly.
We also keep the weight function~\weight\ general;
later in the numerical part of the paper,
we exemplify our approach for $\weightAE=\log[\argument]$ and discuss the weight function more generally.
The additional factor~$\argument$ in the objective function could be absorbed into~\weightE,
but keeping it separate turns out to be more convenient in our calculations---see the factor $1/\argumentE_{\indexelement}$ in the score function below. 

The score functions  $\scorefunction[\argument;\complementaryparameter,\modifiedvariogramL]=\nabla_{\argument} \log \densityF{\argument;\complementaryparameter,\modifiedvariogramL}$ of the model
are comparably simple in our setup:
\begin{lemma}[Score functions]
	\label{scorefunctions}
	In the general model of Section~\ref{generalizedmodel},
	the elements of the score function $\scorefunction=(\scorefunctionE_1,\dots,\scorefunctionE_{\dimfull})$ with $\scorefunction\equiv\scorefunction[\argument;\complementaryparameter,\modifiedvariogramL]$ are
	\begin{equation*}
		\scorefunctionE_{\indexelement}[\argument;\complementaryparameter,\modifiedvariogramL]~=~ \frac{\complementaryparameterE_{\indexelement}-1-\bigl((\modifiedvariogramL+\modifiedvariogramL\tp\varioext)\log[\argument]\bigr)_{\indexelement}}{\argumentE_{\indexelement}}\,.
	\end{equation*}
\end{lemma}
\noindent Hence, the score functions are linear in the parameters.

The formulation of $\objectivefunction$ in \eqref{scalingscorematching} is not ready for estimation yet,
because it involves the ``true'' score function $\scorefunction$ in an upleasant way.
In the following,
we replace that formulation with one that is readily amenable to estimation.
For this, we need to make some mild assumptions on the behavior of the density on the boundary of~\maxargumentS:
\begin{assumption}[Regularity of data-generating density and score]
	\label{ass:dgd}
	Assume that the data-generating density~$\empiricaldensity$ is continuously differentiable on~$\maxargumentS$ and
	\begin{itemize}
		\item $\lim_{\argumentE_{\indexelement}\to \infty} \empiricaldensityF{\argument} (\weightEF{\argumentE_{\indexelement}})^2 \argumentE_{\indexelement} \log[\argumentE_{\indexelement}] = 0 $ for all $\indexelement\in\{1,\dots,\dimfull\}$ and $\argumentE_{-\indexelement}$\,;
		\item $\lim_{\norm{\argument}\to 1+} \empiricaldensityF{\argument} (\weightEF{\argumentE_{\indexelement}})^2 \argumentE_{\indexelement} (1+ \log[\argumentE_{\indexelement}] ) = 0 $ for all $\indexelement\in\{1,\dots,\dimfull\}$\,.
	\end{itemize}
	Furthermore, we need to assume the finiteness of the following moments:
	\begin{itemize}
		\item $\mathbb{E}_{\empiricaldensity} \left(
		\normtwoB{\scorefunction[\argument ; \complementaryparameter,\modifiedvariogramL] \otimes\argument\otimes\weightA}^{2} \right) < \infty $\,;		
		\item $\mathbb{E}_{\empiricaldensity} \left(
		\normoneB{\scorefunction[\argument ; \complementaryparameter,\modifiedvariogramL] \otimes(\argument)^2\otimes(\weightA)^2} \right) < \infty $\,,
	\end{itemize}
	compare to Assumptions (A0.1) and (A0.2) in \cite{Yu2020}.
\end{assumption}
\noindent 
The assumptions are very mild indeed;
in particular,
the assumptions hold for data from a \hr model and $\weightAE=O(\argument^t)$ for a finite~$t$.

For convenience,
we also introduce three functions of the data.
Given a datum~$\argument\in\R^{\dimfull}$,
we define the two vectors
\begin{align*}
	\datafunctionA~&\deq~\bigl(\weightEF{\argumentE_{1}},\dots,\weightEF{\argumentE_{\dimfull}}\bigr)\tp\,\in\,\R^{\dimfull}\\
	\datafunctionPA~&\deq~\bigl((2\weightEF{\argumentE_{1}})^2+4\argumentE_{1}\weightE'[\argumentE_{1}]\weightEF{\argumentE_{1}},\dots,2(\weightEF{\argumentE_{\dimfull}})^2+4\argumentE_{\dimfull}\weightE'[\argumentE_{\dimfull}]\weightEF{\argumentE_{\dimfull}}\bigr)\tp\,\in\,\R^{\dimfull}
\end{align*}
and a diagonal matrix~$\datamatrixA\in\R^{\dimfull\times\dimfull}$ via $(\datamatrix[\argument])_{\indexelement\indexelement} \deq 2(\weightEF{\argumentE_{\indexelement}})^2$ and $(\datamatrix[\argument])_{\indexelement\indexelement} \deq 0$ for $\indexelement\neq\indexelementP$.
We then get the following result:

\begin{theorem}[Score-matching objective]\label{scorematchinggeneral}
	Under Assumption~\ref{ass:dgd},
	the function in (\ref{scalingscorematching}) can be expressed in the form
	\begin{equation*}
		\objectivefunction[\complementaryparameter,\modifiedvariogramL]~=~\frac{1}{2}\int_{\maxargumentS}\empiricaldensityF{\argument}\, \objectivefunctionI[\complementaryparameter,\modifiedvariogramL,\argument]\, \mathrm{d} \argument+\operatorname{const.} ~<~\infty
	\end{equation*}
	where $\operatorname{const.}$ a finite term that is independent of the parameters and
	\begin{multline*}
		\objectivefunctionI[\complementaryparameter,\modifiedvariogramL,\argument]~\deq~\normtwosB{\bigl(\complementaryparameter-\mones-(\modifiedvariogramL+\modifiedvariogramL\tp\varioext)\log[\argument]\bigr)\otimes\datafunctionA}\\
		+ \bigl(\complementaryparameter-\mones-(\modifiedvariogramL+\modifiedvariogramL\tp\varioext)\log[\argument]\bigr)\tp\datafunctionPA\\
		-\tr\Bigl[\bigl(\modifiedvariogramL+\modifiedvariogramL\tp\varioext\bigr)\datamatrixA\Bigr] \,.
	\end{multline*}
\end{theorem}
\noindent 
It is clear that the function $(\complementaryparameter,\modifiedvariogramL)\mapsto\objectivefunctionI[\complementaryparameter,\modifiedvariogramL,\argument]$ is convex for every $\argument\in\mathbb{R}^{\dimfull}$.

Theorem~\ref{scorematchinggeneral} readily leads to an estimator for the underlying parameters~$\complementaryparameter$ and~$\modifiedvariogramL$:
we can simply minimize the empirical version of~$\objectivefunction$.
\begin{definition}[Basic estimator]
	\label{unreg}
	Given data $\argument_1,\dots,\argument_{\numberofobservations}\in\mathbb{R}^{\dimfull}$,
	our basic estimator is any solution of
	\begin{equation*}
		(\complementaryparameterest,\modifiedvariogramLest)~\in~\argmin_{\complementaryparameter\in\locationS,\modifiedvariogramL\in\variogrammatricesS}\biggl\{\sum\nolimits_{\indexsample=1}^{\numberofobservations}\objectivefunctionI[\complementaryparameter,\modifiedvariogramL,\argumenti]\biggr\}\,.
	\end{equation*}
\end{definition}
\noindent
The estimator's objective function is clearly convex and computationally attractive more generally.
Moreover, the asymptotic convergence of the estimator is established in the following section.

But more interesting to us are high-dimensional versions of the basic estimator.
In brief, 
high-dimensional statistics is about models that have many parameters \citep{Lederer2022}.
The flexibility of such models is not only interesting in extremes;
instead, 
such models, and high-dimensional statistics more generally, have already become a cornerstone of statistical data analysis.
However, the high dimensionality brings about three challenges:
(i)~potential overfitting; (ii)~potentially harder interpretations; and (iii)~computational complexity.
High-dimensional statistics addresses these challenges through regularization,
i.e., the introduction of prior functions that formalize additional information or intuition about the problem, similar to Bayesian approaches. 
Prior functions that are convex and induce sparsity have turned out particularly effective in the past.

Our framework lends itself to these ideas especially well
because the objective function of Definition~\ref{unreg} can readily be complemented with a prior function.
We then get:
\begin{definition}[Regularized estimator]
	\label{regscore}
	Given data $\argument_1,\dots,\argument_{\numberofobservations}\in\mathbb{R}^{\dimfull}$, a prior term $\prior\,:\,\R^{\dimfull}\times\R^{\dimfull\times\dimfull}\to[0,\infty]$, and a tuning parameter $\tuningparameter\in[0,\infty)$,
	our regularized estimator is any solution of
	\begin{equation*}
		(\complementaryparameterest,\modifiedvariogramLest)~\in~\argmin_{\complementaryparameter\in\locationS,\modifiedvariogramL\in\variogrammatricesS}\biggl\{\sum\nolimits_{\indexsample=1}^{\numberofobservations}\objectivefunctionI[\complementaryparameter,\modifiedvariogramL,\argumenti]+\sqrt{\nbrsamples}\,\tuningparameter\,\priorF{\complementaryparameter,\modifiedvariogramL}\biggr\}\,.
	\end{equation*}
	
\end{definition}
\noindent
An example for the prior is the $\ell_1$-function $\priorF{\complementaryparameter,\modifiedvariogramL}\deq\sum_{\indexelement=1}^{\dimfull}\abs{\complementaryparameterE_j}+\sum_{\indexelement,\indexelementP=1}^{\dimfull}\abs{\modifiedvariogramLE_{\indexelement\indexelementP}}$,
which has become particularly popular in statistics and machine learning in view of its sparsity-inducing properties.
We illustrate our estimator with the $\ell_1$-prior in the numerical sections,
but we keep our methodological and theoretical considerations more general.

Before that, however, let us come back once more to the normalization constant~$\partitiontrad$ in the Hüsler-Reiss model and our ``normalization constant''~\partition.
In the light of rapid developments in diffusion models~\citep{song2020score,ho2020denoising}, 
score matching seems an interesting and timely approach to extremes in any case.
However, one might ask whether one possibly could apply a more classical approach such as maximum likelihood.
Indeed, there is an explicit expression for~$\partitiontrad$ in the Hüsler-Reiss model~\citep[Proposition 7]{2019Ho} with $\norm{\,\cdot\,}=\normsup{\,\cdot\,}$.
But this explicit expression is still very hard to compute (for example, it involves the inversion of very large matrices),
and it is not clear how this extends beyond the special case of $\normsup{\,\cdot\,}$.
The same holds true for an interesting generalization of the \hr{} model \citep[Section~7.2]{2019Kiriliouk}, which actually comes close to our model.
In our model,  $\densityF{\argument;\complementaryparameter,\modifiedvariogramL}$ might not even be integrable in the first place.

\section{Statistical Theory}
	\label{sec:theory}

This section establishes statistical theories both for the unregularized and the regularized estimator. Some theories for score matching have already been developed but these theories neither cover our version of score matching nor take the involved boundary conditions of the \hr model into account. Hence, the goal of this section is to provide sensible mathematical support for our approach.

\paragraph*{Low-dimensional theory}
We start with low-dimensional settings, which can be approached with the plain estimator of Definition~\ref{unreg}. For simplicity, we assume that the data $\argument_1,\dots,\argument_{\numberofobservations}$ are independent realizations from the domain of attraction of a standard \hr model with variogram matrix~$\variogram\in\variogramS$ as described in Section~\ref{HueslerReiss}.

\begin{theorem}[Low-dimensional theory]
	\label{unregtheory}
	Let $\widehat\variogram\equiv\widehat\variogram[\complementaryparameterest,\modifiedvariogramLest]$ be the variogram that results from the unregularized score-matching estimator given in Definition~\ref{unreg}---see also Lemma~\ref{reparametrization}.
	Then,
	under suitable conditions on the convergence to the \hr model,
	$\widehat\variogram$ converges to $\variogram$ almost surely.
\end{theorem}

\noindent
It should also be straightforward to establish inferential results or include model misspecifications along established lines (see \citet{White1982}, for example),
but our main focus is the high-dimensional theory established in the next section:
Theorem~\ref{unregtheory} should merely hint at the fact that our methodology can be of interest also in low-dimensional settings.
Observe also that the almost-sure convergence implies that for sufficiently many samples,
the signs of the eigenvalues and the rank of the estimated matrices are correct,
meaning that the estimator~$\variogramest$ (and, equivalently, $\modifiedvariogramest$) together with the corresponding, uniquely defined $\complementaryparameter \equiv \complementaryparameter[\variogramest]$ indeed yields a valid \hr distribution.

\paragraph*{High-dimensional theory}
We now turn to high-dimensional settings,
which are our main focus,
and which can be approached with the regularized estimator of Definition~\ref{regscore}.
We consider general parameters~\complementaryparameter\ and~$\modifiedvariogramL$ unless stated otherwise.
However, we limit ourselves to the model itself (rather than the domain of attraction) to obtain finite-sample guarantees.
We lift this restriction again in the numerical parts (see Section~\ref{sec:attraction}).

We first specify our framework and make assumptions in line with standard theories in high-dimensional statistics.
We consider a general class of prior functions that satisfy the following:
\begin{assumption}[Prior function]
	\label{ass:sepa}
	We assume that $ \priorF{\boldsymbol{a},\boldsymbol{A}}=0$ if and only if both $\boldsymbol{a}=\boldsymbol{0}_{\nbrdim}$ and $\boldsymbol{A}=\boldsymbol{0}_{\nbrdim\times\nbrdim}$
	as well as $\priorF{b\boldsymbol{a},b\boldsymbol{A}}=b\,\priorF{\boldsymbol{a},\boldsymbol{A}}$ for all $\boldsymbol{a}\in\R^{\nbrdim}$, $\boldsymbol{A}\in\R^{\nbrdim\times\nbrdim}$, $b\geq 0$.
	Furthermore, we assume that for every index sets $\sparsityV\subset\{1,\dots,\nbrdim\}$ and $\sparsityM\subset\{1,\dots,\nbrdim\}^2$,
	\begin{equation*}
		\priorF{\boldsymbol{a},\boldsymbol{A}}~=~\priorF{\boldsymbol{a}\sparsityVU,\boldsymbol{A}\sparsityMU}+\priorF{\boldsymbol{a}\sparsityVUC,\boldsymbol{A}\sparsityMUC}
	\end{equation*}
	and
	\begin{equation*}
		\priorF{\boldsymbol{a}\sparsityVU,\boldsymbol{A}\sparsityMU}~\leq~\constprior\sqrt{\abs{\sparsityV}}\normtwo{\boldsymbol{a}\sparsityVU}+\constprior\sqrt{\abs{\sparsityM}}\normf{\boldsymbol{A}\sparsityMU}
	\end{equation*}
	for a fixed $\constprior\in(0,\infty)$ and
	for all $\boldsymbol{a}\in\R^{\nbrdim}$ and $\boldsymbol{A}\in\R^{\nbrdim\times\nbrdim}$.
\end{assumption}
\noindent
The first two parts of this assumption ensure positive definiteness and homogeneity---compare to \citet[Equations~(2) and~(3)]{Zhuang2018}.
The third part ensures that the regularizer is decomposable---compare to \citet[Definition~6.4.1]{Lederer2022}---and ``almost'' bounded by an $\ell_2$-norm.
For $\ell_1$-regularization, for example, Assumption~\ref{ass:sepa} is clearly satisfied with~$\constprior=1$.

We then assume that the data allow for parameter estimation in the first place:
\begin{assumption}[Restricted eigenvalue]
	\label{ass:re}
	Consider any fixed $\complementaryparameter\in\R^{\nbrdim}$ and $\modifiedvariogram\in\R^{\nbrdim\times\nbrdim}$.
	We assume that there is a constant $\restrictedeigenvalue\in(0,\infty)$ such that 
	\begin{equation*}
		\sum\nolimits_{\indexsample=1}^{\numberofobservations}\normtwosB{\bigl(\Delta_{\complementaryparameter}-\boldsymbol{\Delta}_{\modifiedvariogram}\log[\argumenti]\bigr)\otimes\datafunctionAi}~\geq~\restrictedeigenvalue\,\numberofobservations\normtwos{(\Delta_{\complementaryparameter})\sparsityVU}+\restrictedeigenvalue\,\numberofobservations\normfs{(\boldsymbol{\Delta}_{\modifiedvariogram})\sparsityMU}
	\end{equation*}
	for all $ ({\Delta}_{\complementaryparameter},{\Delta}_{\modifiedvariogram})\in\mathcal{C}[\sparsityV,\sparsityM]$ and for $\sparsityV\deq\{\indexelement\,:\,\complementaryparameterE_{\indexelement}\neq 0\}$, $\sparsityM\deq\{(\indexelement,\indexelementP)\,:\,\modifiedvariogramE_{\indexelement\indexelementP}\neq 0\}$.
\end{assumption}
\noindent
This assumption aligns with standard assumptions in high-dimensional statistics~\citep{vandeGeer2009}.

We finally require the tuning parameter to be sufficiently large to overrule the noise.
In our framework, the effective noise is \citep{LedererVogt2021}
\begin{equation*}
	\tuningparameterOA~\deq~\priorDFBBB{\sum\nolimits_{\indexsample=1}^{\numberofobservations}\nabla\objectivefunctionI[\complementaryparameter,\modifiedvariogram,\argumenti]}\,,
\end{equation*}
where~\priorD\ is the H\"older-dual of the prior term~$\prior$ according to~\citet[Page~4]{Zhuang2018} (with the natural inner product on the parameter space).
As usual,
we then assume that the effective noise is dominated by the tuning parameter:
\begin{assumption}[Tuning parameter]
	\label{ass:tp}
	Consider any fixed $\complementaryparameter\in\R^{\nbrdim}$ and $\modifiedvariogram\in\R^{\nbrdim\times\nbrdim}$.
	We assume that
	$\tuningparameter~\geq~2\tuningparameterOA$.
\end{assumption}
\noindent 
Broadly speaking, this assumption ensures that we avoid overfitting.

Our main theoretical result is then:
\begin{theorem}[High-dimensional theory]
	\label{regtheory}
	Under Assumptions~\ref{ass:sepa}--\ref{ass:tp},
	it holds that
	\begin{equation*}
		\normtwos{\complementaryparameterest-\complementaryparameter}+\normfs{\modifiedvariogramest-\modifiedvariogram}~\leq\,c\times\frac{\constpriors\bigl(\abs{\sparsityV}+\abs{\sparsityM}\bigr)\tuningparameter^2}{\restrictedeigenvalues\nbrsamples}\,,
	\end{equation*}
	where $c$ is a numerical constant and $(\complementaryparameterest,\modifiedvariogramLest)$ is our regularized score-matching estimator in Definition~\ref{regscore}.
\end{theorem}
\noindent
This is an oracle inequality for the parameters of the generalized \hr model.
The tuning parameter appears as~$\tuningparameter^2$ (rather than $\tuningparameter$)---such oracle inequalities are often called ``power-two'' or ``fast-rate'' bound.
Note that the oracle inequality itself is deterministic;
the randomness is captured by Assumption~\ref{ass:tp}.
Oracle inequalities like the one in Theorem~\ref{regtheory} have been established for many high-dimensional frameworks but not yet for extreme-value setups.

The theorem also implies ``sparsistency:''
\begin{corollary}[Sparsistency]
	Consider the setup of Theorem~\ref{regtheory}, 
	and define the thresholded estimators $\widetilde{\boldsymbol{\mu}},\widetilde{\boldsymbol{\Lambda}}$ through their entries 
	$$\widetilde{\mu}_j\deq\widehat{\mu}_j\indicator{\widehat{\mu}_{j}>t} \quad \text{and} \quad \widetilde{\Lambda}_{jk}\deq\widehat{\Lambda}_{jk}\indicator{\widehat{\Lambda}_{jk}>t}$$
	with threshold $t\deq  \sqrt{c\,\constpriors(\abs{\sparsityV}+2\abs{\mathcal{S}_{\boldsymbol{\Lambda}}}+d)\tuningparameter^2/(\restrictedeigenvalues\nbrsamples)}$.
	Then, assuming that the true model parameters satisfy $\mu_{j}\indicator{\abs{\mu_{j}}\leq 2t}=\Lambda_{jk}\indicator{\abs{\Lambda_{jk}}\leq 2t}=0$ for all $j,k$,
	it holds that
	\begin{equation*}
		\{j~:~\widetilde{\mu}_{j}\,\neq\,0\}~=~ \{j~:~\mu_{j}\,\neq\,0\}~~~~~\text{and}~~~~~\{j,k~:~\widetilde{\Lambda}_{jk}\,\neq\,0\}~=~ \{j,k~:~\Lambda_{jk}\,\neq\,0\}\,.
	\end{equation*}
\end{corollary}
\noindent
Thus, under a standard ``beta-min'' condition,
a properly thresholded version $\widetilde{\boldsymbol{\mu}},\widetilde{\boldsymbol{\Lambda}}$ of the original estimator $\widehat{\boldsymbol{\mu}},\widehat{\boldsymbol{\Lambda}}$ recovers the non-zero patterns of the parameters.
In fact, 
imposing slightly stronger assumptions and applying the primal-dual-witness technique~\citep{2009Wainwright},
one can show that the same holds true for the original, unthresholded estimator as well---but such theoretical subtleties are beyond the scope of this work.
Note that we have formulated the corollary in terms of~$\boldsymbol{\Lambda}$ rather than~$\boldsymbol{\Theta}$:
it turns out that the $\boldsymbol{\Theta}$-based formulation of the general high-dimensional theories is slightly more straightforward,
while the focus on $\boldsymbol{\Lambda}$ is slightly more convenient for the optimization and for the thresholding because, for example, it ensures quite naturally that the thresholded estimators refer to a valid distribution---compare to the low-dimensional theories above.
However, in any case, recall the tight relationship between~$\boldsymbol{\Lambda}$ and~$\boldsymbol{\Theta}$.

\section{Numerical Evidence}
	\label{sec:numerical}

This section demonstrates that our methods work as expected,
modeling multivariate, potentially high-dimensional data at fine resolution.
Moreover,
it highlights the methods' computational efficiency which is
another key feature of our approach.
Indeed, the $\nbrdim+\nbrdim\nbrdimRQ/2\in \mathcal{O}(\nbrdim^2)$ parameters,
which allow for those detailed descriptions of the data,
can be estimated very rapidly:
for example,
one can fit the entire $\ell_1$-regularization path for a model in $\nbrdim=20$ dimensions---which means $210$~parameters---within a second on an ordinary laptop.

\subsection{Setup}
We consider $\ell_1$-regularization,
which is one of the most popular prior functions in statistics and machine learning because it generates a neat, useful type of sparsity. Specifically, we set $\priorF{\complementaryparameter,\modifiedvariogramL}\deq\sum_{\indexelement,\indexelementP=1}^{\dimfull}\abs{\modifiedvariogramLE_{\indexelement\indexelementP}}$;
hence, we do not regularize the intercept nor the diagonal entries of the parameter matrix~$\modifiedvariogram$---as it is also common in Gaussian graphical models (see~\citealp[Equation~(2)]{2021Laszkiewicz}, for example).
Importantly, our methods (see Definition~\ref{regscore}) and theories (see Theorem~\ref{regtheory}) are also amenable to other prior functions.

Our choice for the norm in the definition of~\maxargumentS\ is $\norm{\!\;\cdot\!\;}\deq\normsup{\!\;\cdot\!\;}$.
The sup-norm is standard as it corresponds to the original multivariate Pareto models.
But again, our approach works for other norms as well. As weight function, 
we consider  $\weightAE\deq\log[\argument]$.
This choice of the weight function parallels---at least to some extent---the choices in \citet[Equation~(3)]{Ding2019} and \citet[Equation~(4.21)]{Janofsky2015},
but it is somewhat different from the one in \citet[Equation~(15)]{Fondeville2018}, for example.
The numerical results in this section demonstrate that $\weightAE=\log[\argument]$ works well in our framework.
Nevertheless, we have formulated our methodology and theory for a general weight function~\weight\  to facilitate further research on this topic.
For example, one could try to mimic \citet{Liu2019} and \citet{Yu2020} in their proposal to incorporate the shortest distance to the boundary of the domain (\citet{Liu2019} mainly focus on an overall weight function~$\weightE_1=\dots=\weightE_{\dimfull}$,
whereas \citet{Yu2020} tailor their weight functions to the individual components of the input).
On the one hand, neither \citet{Liu2019} nor \citet{Yu2020} seem to fit our framework directly:
for example, \citet{Yu2020} would lead to $w_j[\argumenthigh]\deq\argumenthigh$ if $\normsup{\argument_{-\indexelement}}>1$ and $w_j[\argumenthigh]\deq\argumenthigh-1$ if $\normsup{\argument_{-\indexelement}}\leq 1$,
which means that the weight functions would not be continuous.
On the other hand,
the weight functions proposed in \citet{Liu2019} and \citet{Yu2020} are similar to ours at the crucial points $\argumenthigh\in\{0, 1\}$ and when $\argumenthigh\to\infty$.

\subsection{Algorithm}
Our basic estimator of Definition~\ref{unreg} is computationally attractive:
the objective function is smooth and convex and, therefore, amenable to standard gradient-based optimization.
The same is true for our regularized estimator of Definition~\ref{regscore} with typical prior functions.
We opt for coordinate descent, 
which allows us to precompute several data-dependent terms before the actual optimization.
As initial values for the optimization,
we chose $\complementaryparameter = \boldsymbol{0}_{\nbrdim}$ and $\modifiedvariogramL = \boldsymbol{0}_{\nbrdim \times \nbrdim}$,
but we did not find a crucial dependence on this choice.
The tuning-parameter path of the regularized estimator is computed as usual:
we start with the largest tuning parameter and then decrease the tuning parameter step-by-step, using the previous results as warm starts each time.
The implementation can be found on our GitHub page \href{https://github.com/LedererLab/HDExtremes}{github.com/LedererLab/HDExtremes}.

We show below that our approach already makes for a very scalable pipeline.
But the framework is also amenable to many additional techniques from convex optimization. In big data applications, for example, where memory can become an issue,
one could easily apply minibatching---both in the actual optimization and the precomputations.
	
\subsection{Synthetic Data}

We first consider synthetic data considering both the case data stemming from the exact limiting model, i.e., the \hr Pareto model, and data from the domain of attraction, i.e., realizations of a random vector $\rv$ such that Equation \eqref{eq:lim-pot} holds with the exponent measure $\mu$ being of \hr type. 

\subsubsection{\hr Pareto data}
We generate $\nbrsamples$ realizations of a $\nbrdim$-dimensional H\"usler--Reiss distribution with a variogram matrix~$\variogram$ that has entries $\variogramE_{ij} \deq \frac{1}{\sqrt{\nbrdim}} |i-j|$ for $i,j\,\in\,\{1,\ldots,\nbrdim\}$. In other words,
$\variogram$ is the variogram matrix of a standard Brownian motion evaluated at times $0, 1/\sqrt{\nbrdim}, \ldots, \sqrt{\nbrdim}-1/\sqrt{\nbrdim}$. As a consequence of the Markov property of the Brownian motion, all the precision matrices
$\modifiedvariogram_{-m,-m} = (\covariance[m])^{-1}$, $m\in\{1,\ldots,\nbrdim\}$, are tridiagonal matrices and, consequently, the parameter matrix~$\modifiedvariogramL$, which is given by
\begin{equation*}
	\modifiedvariogramL ~=~ \begin{pmatrix} 
		0 & -\sqrt{\nbrdim} &   &  \\
		&  0 & -\sqrt{\nbrdim} &       &   \\
		&  & \ddots &  \ddots \\
		&    &        & 0    & -\sqrt{\nbrdim} \\
		&    &    &         & 0
	\end{pmatrix}\,,
\end{equation*}
i.e., $\modifiedvariogramL$ is a sparse matrix with only $\nbrdim-1$ (out of potentially $\nbrdim (\nbrdim-1)/2$) non-zero entries. 
Note that the matrix~$\variogram$ is chosen such that both $\normsup{\variogram} \in \mathcal{O}(\sqrt{\nbrdim})$ and $\normsup{\modifiedvariogramL} \in \mathcal{O}(\sqrt{\nbrdim})$. 

We test both our unregularized (Definition~\ref{unreg}) and regularized estimators (Definition~\ref{regscore}).
Table~\ref{tab:simu_pareto} states
\begin{itemize}
	\item the computing time $t_{\mathrm{pre}}$ in seconds for the precalculations on a standard laptop (Intel~Core~i7, 1.80\,GHz $\times$ 8);
	\item the computing time $t_{\mathrm{opt}}$ in seconds for the calculation of the tuning parameter path of the regularized estimator on the same laptop;  
	\item the root-mean-squared error of $\modifiedvariogramLest$:
	$$ \mathrm{RMSE}_{\modifiedvariogramLest} ~\deq~ \biggl( \frac{2}{\nbrdim (\nbrdim -1)} \sum\nolimits_{1 \leq i < j \leq d}^{\nbrdim} (\widehat \modifiedvariogramLE_{ij} - \modifiedvariogramLE_{ij})^2 \biggr)^{1/2}\;;$$
	\item the relative number of zero-valued entries in the strict upper triangle of $\modifiedvariogramLest$:
	$$ \frac{2}{\nbrdim(\nbrdim -1)} \sum\nolimits_{1 \leq i < j \leq d}^{\nbrdim} \mathbf{1}\{\widehat \modifiedvariogramLE_{ij} = 0\} $$
	\item the root-mean-squared error
	\begin{equation*}
		\mathrm{RMSE}_{\variogramest}~ \deq~ \biggl( \frac{1}{\nbrdim^2} \sum\nolimits_{i=1}^{\nbrdim} \sum\nolimits_{j=1}^{\nbrdim} (\variogramestE_{ij} - \variogramE_{ij})^2 \biggr)^{1/2}
	\end{equation*}
	of the variogram's estimates (see explanations below)\,.
\end{itemize}
More precisely, Table~\ref{tab:simu_pareto} gives those values averaged over $N=100$ simulations in $\nbrdim=20$ and $\nbrdim=80$ dimensions.First, we find that our estimates can recover the true parameters~$\modifiedvariogramL$ very fast;
for example, it takes less than a second to compute the entire tuning-parameter path in $\nbrdim=20$~dimensions (fitting for a single tuning parameter would then be even faster). 
Note that this time does not depend on~$\nbrsamples$, while the time for the precalculations scales roughly linearly in~$\nbrsamples$. 

The number of zero-valued entries is as expected: the larger the tuning parameter, the more zero-valued entries. Next, note that the estimates $\widehat \modifiedvariogramL$ of~$\modifiedvariogramL$ also entail estimates~$\variogramest$ of the variogram~$\variogram$ via the relations~\eqref{eq:vario2covar} and \eqref{eq:def-modifiedvariogram}.
Quite strikingly, even though these calculations seem to rely on the index~$\indexhigh \in \{1,\ldots,\nbrdim\}$, we obtain the same estimates $\variogramest$ (within numerical precision) for all~$\indexhigh$. This result matches the theoretical properties of the H\"usler--Reiss model and, therefore, illustrates the conceptual coherence of our approach.
(Estimates such as those in~\citet{Engelke2022}, in strong contrast, change with~$\indexhigh$.)
Finally, the accuracy of the parameter estimation matches our expectations.
Both $\mathrm{RMSE}_{\modifiedvariogramest}$ and $\mathrm{RMSE}_{\variogramest}$ tend to zero as $\nbrsamples \to \infty$ at a rate of about~$1/\sqrt{\nbrsamples}$.

\begin{table}
	\caption{\label{tab:simu_pareto}Results for the H\"usler--Reiss Pareto model with $\nbrdim=20$ (top) and $\nbrdim=80$ (bottom); numbers are mean values of $N=100$ simulations with standard deviations in brackets}
	\begin{tabular}{|c||c|c|c|c|c|c|c|}
		\hline
		\rowcolor{lightgray} \multicolumn{8}{|c|}{ \textbf{input size $\mathbf{\nbrdim=20}$; model dimension $\mathbf{\nbrdim(\nbrdim+1)/2=210}$}}\\
		\hline  \hline \rowcolor{white}
		\multicolumn{8}{|c|}{\textbf{sample size $\mathbf{\nbrsamples=500}$}} \\ \hline
		time $t_{\mathrm{pre}}$ (s) & \multicolumn{7}{c|}{$0.03$}\\ \hline        
		time $t_{\mathrm{opt}}$ (s) & \multicolumn{7}{c|}{$0.85 (0.11)$}\\ \hline 
		$\tuningparameter \cdot \sqrt{\nbrsamples / \log[\nbrdim]}$ & $2000 $ & $200 $ 
		& $20$ & $2$ & $0.2$ & $0.02$ & $0$ \\ \hline
		ratio of zeros & $65.8\%$ & $37.6\%$ & \,$3.6\%$ & \,$0.5\%$ & \,$0.0\%$ & \,$0.0\%$ & \,$0.0\%$ \\
		in parameters & $~~(2.6\%)$ & $~~(4.1\%)$ & $~(1.4\%)$ & $~(0.5\%)$ & $~(0.1\%)$ & $~(0.1\%)$ & $~(0.0\%)$ \\ \hline                                     
		$\mathrm{RMSE}_{\modifiedvariogramLest}$ & 0.87 & 0.33 & 0.68 & 0.74 & 0.75 & 0.75 & 0.75 \\
		& (0.051) & (0.044) & (0.067) & (0.070) & (0.071) & (0.071) & (0.071) \\ \hline
		$\mathrm{RMSE}_{\variogramest}$ 
		& 0.51 & 0.28 & 0.25 & 0.25 & 0.25 & 0.25 & 0.25 \\
		& (0.053) & (0.196) & (0.090) & (0.091) & (0.091) & (0.091) & (0.091) \\ \hline \hline
		\multicolumn{8}{|c|}{\textbf{sample size $\mathbf{\nbrsamples=50\,000}$}} \\ \hline
		time $t_{\mathrm{pre}}$ (s) & \multicolumn{7}{c|}{2.38 (0.08)}\\ \hline        
		time $t_{\mathrm{opt}}$ (s) & \multicolumn{7}{c|}{0.83 (0.06)} \\ \hline        
		$\tuningparameter \cdot \sqrt{\nbrsamples / \log[\nbrdim]}$ & $2000 $ & $200 $ 
		& $20$ & $2$ & $0.2$ & $0.02$ & $0$ \\ 	\hline	
		ratio of zeros  & $34.6\%$ & $3.6\%$ & $0.3\%$ & $0.0\%$ & $0.0\%$ & $0.0\%$ & $0.0\%$ \\
		in parameters & $~~(2.9\%)$ & $(1.5\%)$ & $(0.4\%)$ & $(0.2\%)$ & $(0.0\%)$ & $(0.0\%)$ & $(0.0\%)$  \\ \hline  
		$\mathrm{RMSE}_{\modifiedvariogramLest}$ & 0.04 & 0.06 & 0.07 & 0.07 &
		0.07 & 0.07 & 0.07 \\
		& (0.005) & (0.005) & (0.005) & (0.005) & (0.005) & (0.005) & (0.005)  \\ \hline
		$\mathrm{RMSE}_{\variogramest}$ 
		& 0.03 & 0.02 & 0.02 & 0.02 & 0.02 & 0.02 & 0.02 \\
		& (0.013) & (0.009) & (0.009) & (0.009) & (0.009) & (0.009) & (0.009) \\ \hline
		\multicolumn{8}{c}{} \\ \hline
		\rowcolor{lightgray} 
		\multicolumn{8}{|c|}{\textbf{input size $\mathbf{\nbrdim=80}$; model dimension $\mathbf{\nbrdim(\nbrdim+1)/2=3240}$}} \\ \hline \rowcolor{white}
		time $t_{\mathrm{pre}}$ (s) & \multicolumn{7}{c|}{~~1.31 (0.02)}\\ \hline        
		time $t_{\mathrm{opt}}$ (s) & \multicolumn{7}{c|}{339.72 (36.23)} \\ \hline 
		& \multicolumn{7}{c|}{\textbf{sample size $\mathbf{\nbrsamples=500}$}} \\
		$\tuningparameter \cdot \sqrt{\nbrsamples / \log[\nbrdim]}$ & $2000 $ & $200 $ 
		& $20$ & $2$ & $0.2$ & $0.02$ & $0$ \\ \hline
		ratio of zeros  & $86.8\%$ & $20.9\%$ & $1.9\%$ & $0.2\%$ & $0.0\%$ & $0.0\%$ & $0.0\%$ \\
		in parameters & ~~$(0.6\%)$ & ~~$(1.2\%)$ & $(0.3\%)$ & $(0.1\%)$ & $(0.0\%)$ & $(0.0\%)$ & $(0.0\%)$ \\ \hline
		$\mathrm{RMSE}_{\modifiedvariogramLest}$ & 0.45 & 1.22 & 2.08 & 2.19 & 2.20 & 2.20 & 2.20 \\
		& (0.022) & (0.069) & (0.090) & (0.093) & (0.094) & (0.094) & (0.094) \\ \hline
		$\mathrm{RMSE}_{\variogramest}$ 
		& 2.31 & 1.91 & 3.58 & 3.38 & 3.01 & 3.01 & 3.01 \\
		& (0.126) & (0.123) & (12.250) & (8.202) & (4.143) & (4.061) & (4.504) \\ \hline
		\hline 
		& \multicolumn{7}{c|}{\textbf{sample size $\mathbf{\nbrsamples=50\,000}$}} \\ \hline
		time $t_{\mathrm{pre}}$ (s) & \multicolumn{7}{c|}{132.91 (1.59)}\\ \hline        
		time $t_{\mathrm{opt}}$ (s) & \multicolumn{7}{c|}{183.32 (10.47)}\\ \hline
		$\tuningparameter \cdot \sqrt{\nbrsamples / \log[\nbrdim]}$ & $2000 $ & $200 $ 
		& $20$ & $2$ & $0.2$ & $0.02$ & $0$ \\ \hline
		ratio of zeros  & $20.3\%$ & $2.0\%$ & $0.2\%$ & $0.0\%$ & $0.0\%$ & $0.0\%$ & $0.0\%$ \\
		in parameters & $~~(0.8\%)$ & $(0.2\%)$ & $(0.1\%)$ & $(0.0\%)$ & $(0.0\%)$ & $(0.0\%)$ & $(0.0\%)$ \\ \hline  
		$\mathrm{RMSE}_{\modifiedvariogramLest}$ & 0.09 & 0.13 & 0.13 & 0.13 & 0.13 & 0.13 & 0.13 \\
		& (0.002) & (0.002) & (0.003) & (0.003) & (0.003) & (0.003) & (0.003) \\ \hline
		$\mathrm{RMSE}_{\variogramest}$ 
		& 0.11 & 0.11 & 0.11 & 0.11 & 0.11 & 0.11 & 0.11 \\
		& (0.057) & (0.056) & (0.056) & (0.056) & (0.056) & (0.056) & (0.056) \\ \hline
	\end{tabular}
\end{table}

\subsubsection{Data from the domain of attraction of the \hr Pareto model}
\label{sec:attraction}
Using the extremal functions approach \citep{DEO16}, we now generate $\nbrsamples$ realizations of a $\nbrdim$-dimensional max-stable H\"usler--Reiss distribution with unit Fr\'echet margins, 
which means that the data do not satisfy Equation \eqref{eq:eq-pot} but only the limiting relation in Equation \eqref{eq:lim-pot}. In order to obtain data that approximately follow a \hr Pareto distribution,
we choose a high threshold $u$ and consider $u^{-1}\rv$ for those realizations $\rv$ that satisfy $\norm{\rv}_\infty > u$.

Here, we take the same variogram matrix as in the previous subsection for $\nbrdim=20$ and  $u$ as the $95$-percentile of the marginal unit Fr\'echet margins. Thus, $u$ is exceeded for roughly $13.5\,\%$ of all realizations. Consequently, in order to obtain comparable effective samples, the numbers for $\nbrsamples$ in Table \ref{tab:simu_pareto} are multiplied by a factor of $7$, i.e., we consider the cases
$\nbrsamples=3500$ and $\nbrsamples=350\,000$, respectively. Furthermore, for the sake of comparability, the tuning parameter $\tuningparameter$ is defined in terms of the random number $\nbrsamples_u$ of threshold exceedances rather than the original sample size $\nbrsamples$.

\begin{table}
	\caption{\label{tab:simu_20_maxstable}Results for the max-stable H\"usler--Reiss model with $\nbrdim=20$, i.e., $210$ parameters; numbers are mean values of $N=100$ simulations with standard deviations in brackets}
	
	\begin{tabular}{|c||c|c|c|c|c|c|c|}
		\hline 
		\rowcolor{lightgray} \multicolumn{8}{|c|}{ \textbf{input size $\mathbf{\nbrdim=20}$; model dimension $\mathbf{\nbrdim(\nbrdim+1)/2=210}$}}\\
		\hline  \hline \rowcolor{white}
		\multicolumn{8}{|c|}{\textbf{sample size $\mathbf{\nbrsamples=3500}$}} \\ \hline
		time $t_{\mathrm{pre}}$ (s) & \multicolumn{7}{|c|}{0.03}\\ \hline        
		time $t_{\mathrm{opt}}$ (s) & \multicolumn{7}{|c|}{0.54 (0.05)}\\ \hline 
		$\tuningparameter \cdot \sqrt{\nbrsamples / \log[\nbrdim]}$ & $2000 $ & $200 $ 
		& $20$ & $2$ & $0.2$ & $0.02$ & $0$ \\ \hline
		ratio of zeros & $49.5\%$ & $48.8\%$ & $5.1\%$ & $0.6\%$ & $0.0\%$ & $0.0\%$ & $0.0\%$ \\
		in parameters & $~~(4.2\%)$ & $~~(4.1\%)$ & $~(1.6\%)$ & $~(0.5\%)$ & $~(0.1\%)$ & $~(0.1\%)$ & $~(0.0\%)$ \\ \hline                                     
		$\mathrm{RMSE}_{\modifiedvariogramLest}$ & 1.15 & 0.29 & 0.68 & 0.77 & 0.78 & 0.78 & 0.78  \\
		& (0.075) & (0.043) & (0.076) & (0.081) & (0.082) & (0.082) & (0.082) \\ \hline
		$\mathrm{RMSE}_{\variogramest}$ 
		& 0.95 & 0.73 & 0.58 & 0.56 & 0.56 & 0.56 & 0.56 \\
		& (0.262) & (0.085) & (0.097) & (0.099) & (0.099) & (0.099) & (0.099) \\ \hline
		\hline 
		\multicolumn{8}{|c|}{\textbf{sample size $\mathbf{\nbrsamples=3500}$}} \\ \hline	
		time $t_{\mathrm{pre}}$ (s) & \multicolumn{7}{|c|}{2.01 (0.06)}\\ \hline        
		time $t_{\mathrm{opt}}$ (s) & \multicolumn{7}{|c|}{0.45 (0.04)} \\ \hline                                    	
		$\tuningparameter \cdot \sqrt{\nbrsamples / \log[\nbrdim]}$ & $2000 $ & $200 $ 
		& $20$ & $2$ & $0.2$ & $0.02$ & $0$ \\ \hline
		ratio of zeros & $41.5\%$ & $4.8\%$ & $0.5\%$ & $0.0\%$ & $0.0\%$ &$0.0\%$ & $0.0\%$ \\
		in parameters & $~~(2.7\%)$ & $(1.6\%)$ & $(0.5\%)$ & $(0.2\%)$ & $(0.0\%)$ & $(0.0\%)$ & $(0.0\%)$ \\ \hline  
		$\mathrm{RMSE}_{\modifiedvariogramLest}$ & 0.06 & 0.09 & 0.09 & 0.09 & 0.09 & 0.09 & 0.09 \\
		& (0.005) & (0.006) & (0.006) & (0.006) & (0.006) & (0.006) & (0.006)  \\ \hline
		$\mathrm{RMSE}_{\variogramest}$ 
		& 0.56 & 0.54 & 0.54 & 0.54 & 0.54 & 0.54 & 0.54  \\
		& (0.012) & (0.012) & (0.012) & (0.012) & (0.012) & (0.012) & (0.012)
		\\ \hline
	\end{tabular}
\end{table}

The results are displayed in Table \ref{tab:simu_20_maxstable}. Most of them are very similar to the ones in the \hr Pareto case in Table \ref{tab:simu_pareto}, 
except for  $\mathrm{RMSE}_{\variogramest}$ being much larger in the max-stable case. This observation can be explained by the fact that the \hr Pareto density is only an approximation of the data generating density in this case and that matrix inversion is very sensitive to small deviations.
	
\begin{figure} 
	\includegraphics[height=6cm]{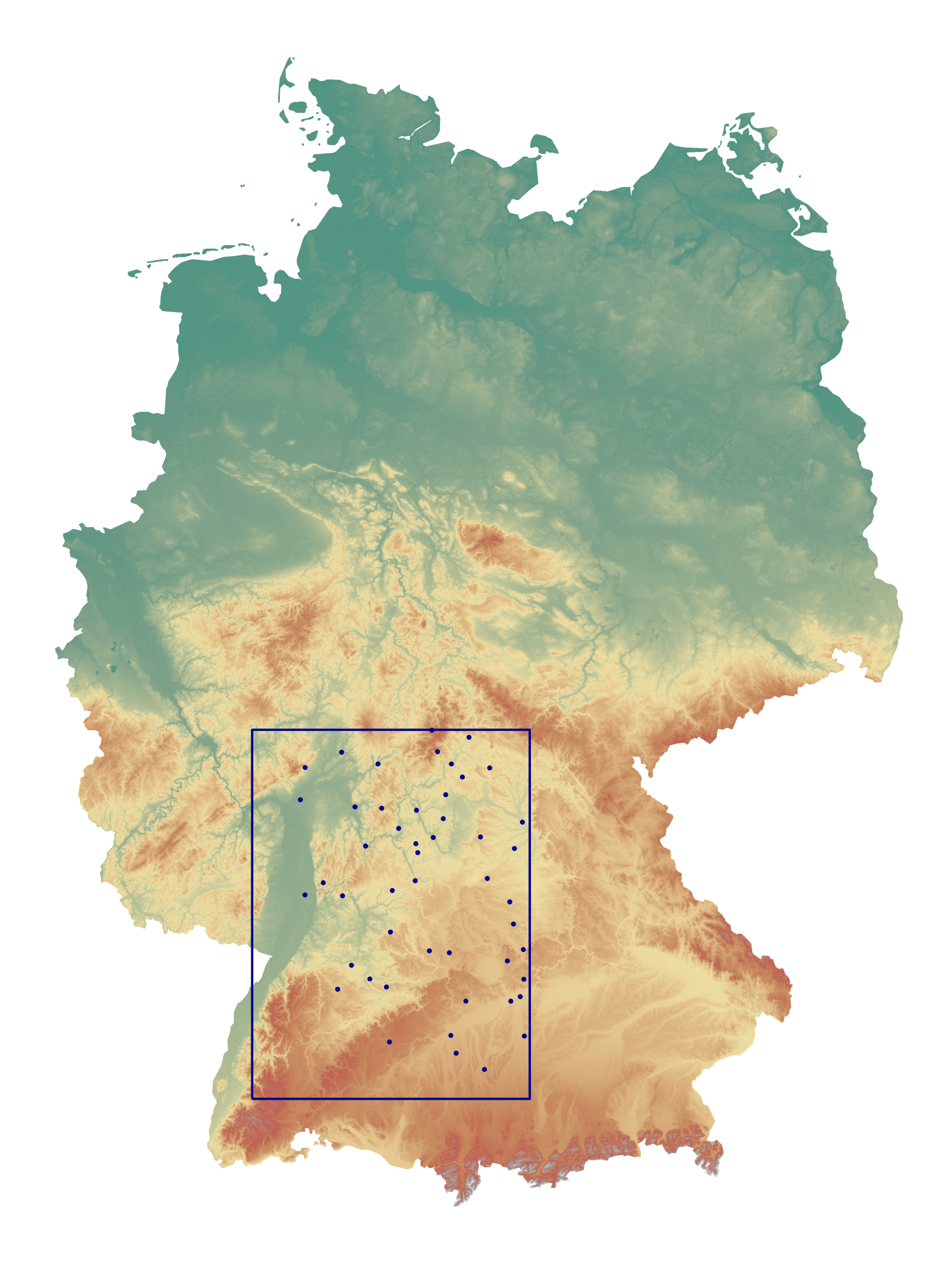}
	\includegraphics[height=6cm]{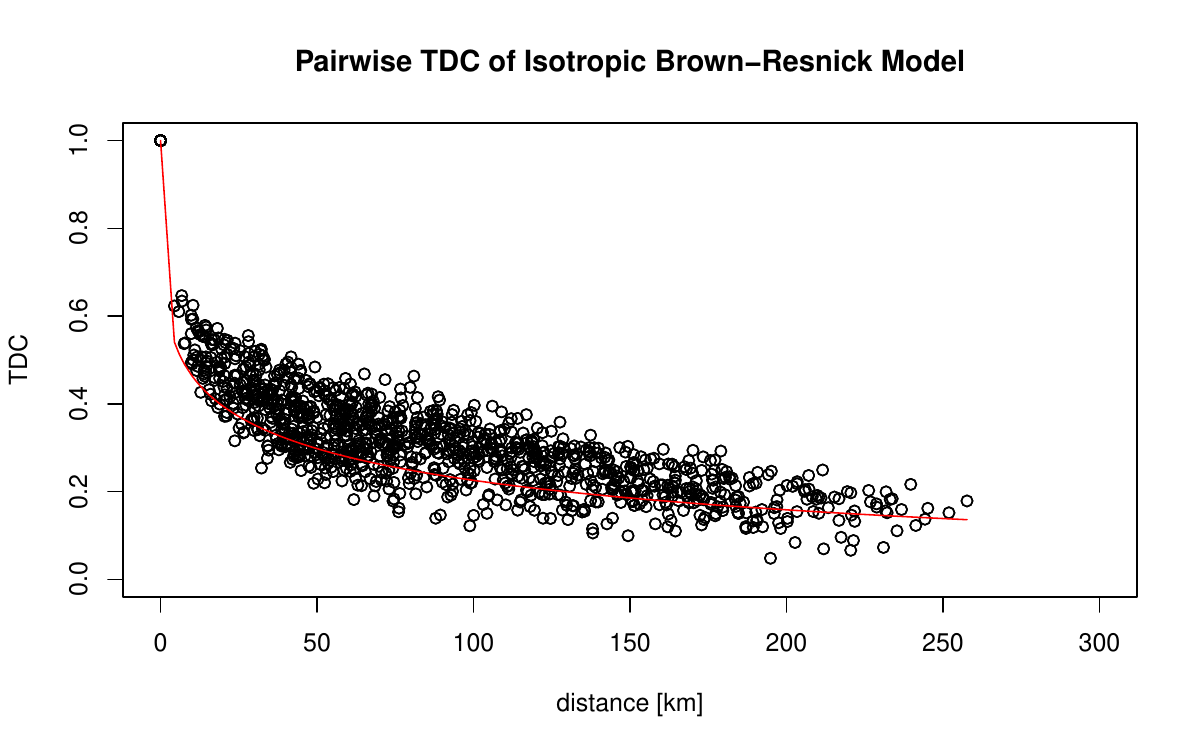}
	\caption{Left: Map of $d=49$ weather stations in the southern part of Germany considered in the study. Right: Empirical pairwise upper tail dependence coefficients $\widehat \chi_{\ell j}$ plotted against the geographical distance between the two stations of each pair. The solid line depicts the theoretical values of the fitted isotropic Brown--Resnick model.} \label{fig:map_BR}
\end{figure}

\subsection{Real Data}

he goal of our estimation pipeline are accurate parameter estimates.
These parameter estimates can then be used directly in downstream analyses, for example, for the estimation of tail dependence coefficients. 
To illustrate such an application, 
we consider precipitation data in Germany. 
We use daily precipitation measurements at $d=49$ weather stations run by the German national meteorological service (DWD) at sites $s_1, \ldots, s_d$ (see the left panel of Figure~\ref{fig:map_BR}) for the summer months (June, July, and August) for 69 consecutive years from 1951 to 2019. Each summer period is split into six disjoint blocks of $15$ days each, and, for each station $j \in \{1,\ldots,d\}$, the maximum over each block is taken.
This yields $n=414$ random vectors $\rvY_1, \ldots, \rvY_n \in \mathbb{R}^{d}$, which are considered independent and identically distributed.

The random vectors are marginally transformed to vectors $\rvZ_1, \ldots, \rvZ_n$ with unit Fr\'echet margins via ranks:
$$ (\rvZE_{i})_j = - \frac{1}{\log\left( \text{rank}((\rvyE_{i})_{j})/(n+1)\right)}, \quad i\in\{1,\ldots,n\}, \ j\in\{1,\ldots,d\}\,.$$
The transformed data are assumed to be independent copies of a max-stable H\"usler--Reiss vector $Z$. Following one of the standard approaches in spatial extremes (see the references in the introduction), a parametric power variogram model
$ \Gamma_{i,j} = c \norm{s_i - s_j}^{\alpha}$, $i,j \in \{1,\ldots,d\}$
with parameters $c>0$ and $\alpha \in (0,2]$ is fitted via maximizing a composite pairwise likelihood \cite{padoan-etal-10} using the \texttt{R} package \texttt{SpatialExtremes} \cite{SpatialExtremes}. Note that, by definition, this model is isotropic, that is, only depends on the geographic distance $\normtwo{s_i - s_j}$.

To study the quality of the fit, 
we investigate the pairwise dependence of $Z_\ell$ and $Z_j$ (or, equivalently, between $Y_\ell$ and $Y_j$) for $\ell,j\in\{1,\dots, d\}$, $\ell\neq j$,  which is typically summarized via the pairwise upper-tail dependence coefficients
$$ \chi_{\ell j} = \lim_{q \uparrow 1} \mathbb{P}\bigl\{Y_\ell > F_\ell^{-1}(q) \mid Y_j > F_j^{-1}(q)\bigr\} = \lim_{u \to \infty}  \mathbb{P}\bigl\{Z_\ell > u \mid Z_j > u\bigr\}\in[0,1]\,, $$
where $F_\ell$ and $F_j$ describe the cumulative distribution functions of $Y_\ell$ and $Y_j$, respectively.
Provided that $\chi_{\ell j}$ exists, it describes the strength of asymptotic dependence between $Z_\ell$ and $Z_j$, where $\chi_{\ell j}=1$ corresponds to perfect asymptotic dependence, while $\chi_{\ell j}=0$ means asymptotic independence.
Provided that $Z$ follows a \hr distribution, the coefficients satisfy
$$ \chi_{\ell j} = 2 \cdot \Big[1 - \Phi\Big(\sqrt{\Gamma_{\ell j}/2}\Big)\Big]\,,$$
where $\Phi$ denotes the c.d.f.~of the standard normal distribution.
In a max-stable setting, these can also be estimated via the relation
$$ \widehat \chi_{\ell j}  = 2 - \frac{1 + 2 \widehat \nu_{\ell j}}{1 - 2 \widehat \nu_{\ell j}}\,,$$
where $\widehat \nu_{\ell j}$ denotes the empirical $F$-madogram \citep{CNP06} given by 
$$ \widehat \nu_{\ell ij} = \frac 1 {2n} \sum\nolimits_{i=1}^n \left| \frac{\text{rank}(Y_{i\ell})}{n+1} - \frac{\text{rank}(Y_{\ell j})}{n+1} \right|\,. $$
We can now plot the spatial distances $\normtwo{s_\ell - s_j}$ against the empirical tail coefficients~$\widehat \chi_{\ell j}$ and compare to the theoretical values of the fitted isotropic \hr model (or, interpreted as a spatial process, Brown--Resnick model). 
The results in the right panel of Figure~\ref{fig:map_BR}
show that the empirical estimates scatter quite widely around the theoretical curve
suggesting that the data-generating process cannot be described solely through the distances among the stations, that is, the plot suggests that the data are not isotropic in space.
Our conclusion is in line with the findings in \cite{forster-oesting-22}, who propose a non-stationary model for similar data in the southern part of Germany.

These results motivate us to fit a more flexible \hr model using  the score-matching estimator proposed in Section~\ref{subsec:scorematching}.
We transform the data marginally via ranks to unit Pareto margins, that is, we consider
$$ X_{ij} = - \frac{n+1}{n + 1 -  \text{rank}(Y_{ij})}, \quad i\in\{1,\ldots,n\}, \ j\in\{1,\ldots,d\}\,.$$
We then take the normalized data $u^{-1} \boldsymbol{X}_{\cdot j}$ conditional on $\normsup{\boldsymbol{X}} > u$ for the threshold $u=10$. 
The resulting $316$~data vectors (out of the original $6279$ daily observations before taking block maxima) are considered as independent realizations of a \hr Pareto model.
This model is then fitted via two different estimators:
\begin{itemize}
	\item the empirical variogram \citep{engelke2022structure} based on exceedances over the threshold $u=10$;
	\item our new regularized score-matching estimator based on the same threshold
	with a small regularization parameter $\tuningparameter=10^{-10}$.
\end{itemize}
\noindent
The empirical variogram is a very simple approach; 
in particular, it does not include any regularization.
However, since the empirical variogram is often used as a starting point in analysis pipelines,
it can serve as a rough benchmark.
To allow for a comparison with that benchmark,
we take a very small tuning parameter for our method. 

\begin{figure} 
	\includegraphics[height=6cm]{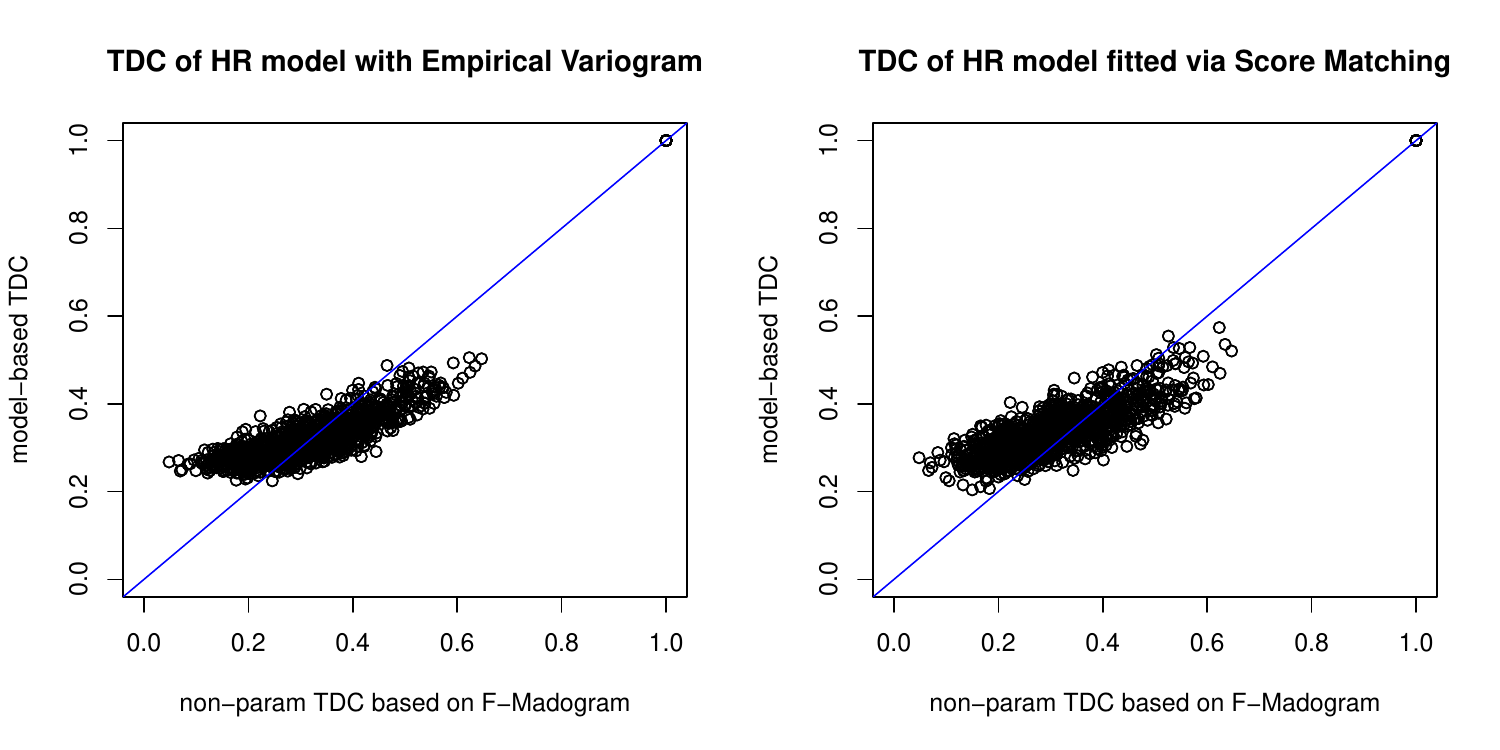}
	\caption{Theoretical pairwise tail dependence coefficients for the \hr models fitted via the empirical variogram (left) and score matching (right) plotted against the empirical estimates.} \label{fig:TDC_HR}
\end{figure}

Figure \ref{fig:TDC_HR} plots the theoretical pairwise dependence coefficients of the fitted models against their empirical counterparts (the same as in Figure~\ref{fig:map_BR}). 
For both estimators, the fitted models  provide a decent fit, 
suggesting that a \hr framework with a sufficiently flexible variogram matrix does indeed make sense.
Moreover, the almost unregularized score-matching estimator is close to the empirical variogram, as expected.
The score-matching estimator can then also be a starting point for more substantial regularization once additional information or intuition becomes available through further studies or domain experts.
Thus,
the application highlights the potential of our fast and flexible \hr pipeline in practice.

\section{Discussion}	\label{sec:discussion}

We introduce modern strategies of machine learning to the field of extremes, namely the concept of regularization, the corresponding theories of high-dimensional statistics,
and the gradient-based algorithms of convex optimization. In this way, we are able to estimate statistically sound models for extremes at unprecedented speed and detail.

A aspect for further study is the weight function \weightE. Our choice satisfies the mathematical requirements and seems to work well in practice. However, it would be interesting to study whether different choices of~\weightE\ might lead to further gains in efficiency in specific applications.

\paragraph*{Acknowledgements}
We thank Anthony Davison, Sebastian Engelke, Micha\"el Lalancette, and Ali Mohaddes for their insightful comments. 

\paragraph*{Funding}
MO gratefully acknowledges funding by Deutsche Forschungsgemeinschaft (DFG, German Research Foundation) under Germany’s Excellence Strategy -- EXC 2075 -- 390740016.

\newpage
\appendix
\begin{center} 
  \Large Supplementary Material: EXTREMES IN HIGH DIMENSIONS: METHODS AND SCALABLE ALGORITHMS
\end{center}
\bigskip

This supplementary material contains two appendices: one for further technical results, one for the proofs.

\section{Further Technical Results}
\label{sec:further}

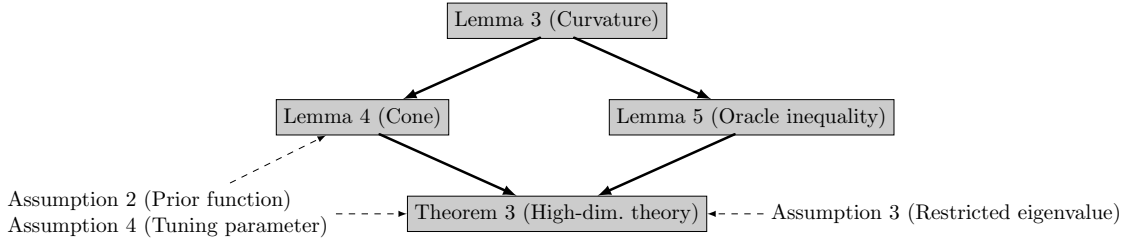
\begin{figure}[h]
	\centering
	\scalebox{0.8}{\begin{tikzpicture}[scale=0.8]
			\node[draw, fill=black!20] (Main) at (0, 0) {Theorem~\ref{regtheory} (High-dim.~theory)};
			\node[draw, fill=black!20] (Curvature) at (0, 4) {Lemma~\ref{res:margin} (Curvature)};
			\node[draw, fill=black!20] (Cone) at (-4, 2) {Lemma~\ref{res:cone} (Cone)};
			\node[draw, fill=black!20] (OE) at (4, 2) {Lemma~\ref{res:oe} (Oracle inequality)};
			
			\node[fill=black!0, align=left] (TP) at (-8, 0) {Assumption~\ref{ass:sepa} (Prior function)\\Assumption~\ref{ass:tp} (Tuning parameter)};
			\node[fill=black!0] (RE) at (8, 0) {Assumption~\ref{ass:re} (Restricted eigenvalue)};
			
			\draw[very thick, -latex, cap=round] (Curvature)+(-10pt, -10pt) -- (Cone);
			\draw[very thick, -latex, cap=round] (Curvature)+(+10pt, -10pt) -- (OE);
			\draw[very thick, -latex, cap=round] (Cone)+(+10pt, -10pt) -- (Main);
			\draw[very thick, -latex, cap=round] (OE)+(-10pt, -10pt) -- (Main);
			
			\draw[-latex, dashed] (TP) -- (Main);
			\draw[-latex, dashed] (TP) -- (Cone);
			\draw[-latex, dashed] (RE) -- (Main);
	\end{tikzpicture}}
	\caption{overview of the high-dimensional theory}
	\label{fig:overviewhd}
\end{figure}

This section contains theoretical results that we use in our proof of the high-dimensional guarantees and that might also be of independent interest.
An overview is provided in Figure~\ref{fig:overviewhd}.

\subsection{Curvature}

Observe first that the objective function~$\objectivefunctionI\equiv\objectivefunctionI[\complementaryparameter,\modifiedvariogramL,\argument]$ depends on its second argument~$\modifiedvariogramL$ only through $\modifiedvariogramL+\modifiedvariogramL\tp\varioext$;
hence, we can write---with some abuse of notation---the values $\objectivefunctionI[\complementaryparameter,\modifiedvariogramL,\argument]$ as $\objectivefunctionI[\complementaryparameter,\modifiedvariogram,\argument]$ with $$\modifiedvariogram\deq\modifiedvariogramL+\modifiedvariogramL\tp\varioext.$$
In other words,
\begin{multline*}
	\objectivefunctionI[\complementaryparameter,\modifiedvariogram,\argument]~=~\normtwosB{\bigl(\complementaryparameter-\mones-\modifiedvariogram\log[\argument]\bigr)\otimes\datafunctionA}
	+\bigl(\complementaryparameter-\mones-\modifiedvariogram\log[\argument]\bigr)\tp\datafunctionPA\\
	-\tr\bigl[\modifiedvariogram\datamatrixA\big]\,.
\end{multline*}
More generally, we use~$\modifiedvariogramL$ and~$\modifiedvariogram$ interchangeably in our notation.
With this in mind, we find the following identity:
\begin{lemma}[Curvature]
	\label{res:margin}
	For every $\complementaryparameter,\complementaryparameter'\in\R^{\dimfull}$ and $\modifiedvariogram,\modifiedvariogram'\in\R^{\dimfull\times\dimfull}$,
	it holds that
	\begin{multline*}
		\sum_{\indexsample=1}^{\numberofobservations}\objectivefunctionI[\complementaryparameter',\modifiedvariogram',\argumenti]~=~\sum_{\indexsample=1}^{\numberofobservations}\objectivefunctionI[\complementaryparameter,\modifiedvariogram,\argumenti]
		+\inprodBBB{\sum_{\indexsample=1}^{\numberofobservations}\nabla\objectivefunctionI[\complementaryparameter,\modifiedvariogram,\argumenti]}{(\Delta_{\complementaryparameter},\boldsymbol{\Delta}_{\modifiedvariogram})}\\
		+\sum_{\indexsample=1}^{\numberofobservations}\normtwosB{\bigl(\Delta_{\complementaryparameter}-\boldsymbol{\Delta}_{\modifiedvariogram}\log[\argumenti]\bigr)\otimes\datafunctionAi}
	\end{multline*}
	and
	\begin{multline*}
		\E_{\argument}\bigl[\objectivefunctionI[\complementaryparameter',\modifiedvariogram',\argument]\bigr]~=~\E_{\argument}\bigl[\objectivefunctionI[\complementaryparameter,\modifiedvariogram,\argument]\bigr]
		+\inprodB{\nabla\E_{\argument}\bigl[\objectivefunctionI[\complementaryparameter,\modifiedvariogram,\argument]\bigr]}{({\Delta}_{\complementaryparameter},\boldsymbol{\Delta}_{\modifiedvariogram})}\\
		+\E_{\argument}\Bigl[\normtwosB{\bigl({\Delta}_{\complementaryparameter}-\boldsymbol{\Delta}_{\modifiedvariogram}\log[\argument]\bigr)\otimes\datafunctionA}\Bigr]\,,
	\end{multline*}
	where ${\Delta}_{\complementaryparameter}\deq\complementaryparameter'-\complementaryparameter$, $\boldsymbol{\Delta}_{\modifiedvariogram}\deq\modifiedvariogram'-\modifiedvariogram$,
	$({\Delta}_{\complementaryparameter},\boldsymbol{\Delta}_{\modifiedvariogram})\in\R^{\dimfull}\times\R^{\dimfull\times\dimfull}$ concatenates the two quantities,
	similarly, $\nabla$ is the gradient with respect to $(\complementaryparameter,\modifiedvariogram)$,
	and $\inprod{\cdot}{\cdot}$ is the canonical inner product on $\R^{\dimfull}\times\R^{\dimfull\times\dimfull}$.
\end{lemma}
\noindent
This identity shows that the objective function has a quadratic curvature,
which highlights the estimator's benign behavior in terms of both statistical theory and numerical computations.

\subsection{Cone}

Given a vector $\boldsymbol{a}\in\R^p$ and a set $\mathcal{S}\subset\{1,\dots,p\}$,
we denote by $\boldsymbol{a}_{\mathcal{S}}\in\R^{|\mathcal{S}|}$ the subvector of~$\boldsymbol{a}$ whose elements have index in~$\mathcal{S}$;
for example,
$((4, 5, 6)\tp)_{\{1,3\}}=(4,6)\tp$.
Similarly, given a matrix $\boldsymbol{A}\in\R^{p\times p}$ and a set $\mathcal{S}\subset\{1,\dots,p\}^2$,
we denote by $\boldsymbol{A}_{\mathcal{S}}\in\R^{|\mathcal{S}|}$ the submatrix of~$\boldsymbol{A}$ whose elements have index in~$\mathcal{S}$.
For every $\boldsymbol{a}\in\R^{\dimfull}$, $\boldsymbol{A}\in\R^{\dimfull\times\dimfull}$, $\sparsityV\subset\{1,\dots,p\}$, and $\sparsityM\subset\{1,\dots,p\}^2$,
we write $(\boldsymbol{a},\boldsymbol{A})\in\mathcal{C}[\sparsityV,\sparsityM]$ if
\begin{equation*}
	3\priorF{\boldsymbol{a}\sparsityVU,\boldsymbol{A}\sparsityMU}~\geq~\priorF{\boldsymbol{a}\sparsityVUC,\boldsymbol{A}\sparsityMUC}\,.
\end{equation*}
The factor~$3$ is arbitrary:
any factor strictly large than~$1$ works if the condition on the tuning parameter in the following lemma is adjusted accordingly.
\begin{lemma}[Cone]
	\label{res:cone}
	Consider any $\complementaryparameter\in\locationS$ and $\modifiedvariogram\in\variogrammatrices$,
	and set $$\modifiedvariogramest\deq\modifiedvariogramLest+\modifiedvariogramLest\tp\varioextest\,,$$
	where $(\complementaryparameterest,\modifiedvariogramLest)$ is our regularized score-matching estimator in Definition~\ref{regscore}.
	Then, under Assumptions~\ref{ass:sepa} and~\ref{ass:tp},
	it holds that
	\begin{equation*}
		({\Delta}_{\complementaryparameter},{\Delta}_{\modifiedvariogram})~\in~\mathcal{C}[\sparsityV,\sparsityM]\,,
	\end{equation*}
	where ${\Delta}_{\complementaryparameter}\deq\complementaryparameter-\complementaryparameterest$,
	${\Delta}_{\modifiedvariogram}\deq\modifiedvariogram-\modifiedvariogramest$,
	and $\sparsityV\deq\{\indexelement\,:\,\complementaryparameterE_{\indexelement}\neq 0\}$, $\sparsityM\deq\{(\indexelement,\indexelementP)\,:\,\modifiedvariogramE_{\indexelement\indexelementP}\neq 0\}$.
\end{lemma}
\noindent
Broadly speaking,
this result ensures that the estimator's difference to any parameter is concentrated on the support of that other parameter---cf.~\citet[Section~6.4]{Lederer2022}.
Again, in the condition on the tuning parameter in Assumption~\ref{ass:tp}, any factor strictly larger than~$1$ is possible as long as the factor in the definition of the cone is adjusted accordingly.

\subsection{Oracle Inequality}
We can now state our first oracle inequality:
\begin{lemma}[Oracle inequality]
	\label{res:oe}
	In the notation of Lemma~\ref{res:cone},
	it holds that
	\begin{equation*}
		\sum_{\indexsample=1}^{\numberofobservations}\normtwosB{\bigl(\Delta_{\complementaryparameter}-\boldsymbol{\Delta}_{\modifiedvariogram}\log[\argumenti]\bigr)\otimes\datafunctionAi}~\leq~\bigl(\tuningparameter+\tuningparameterOA\bigr)\priorF{\Delta_{\complementaryparameter},\boldsymbol{\Delta}_{\modifiedvariogram}}\,.
	\end{equation*}
\end{lemma}
\noindent
This result is a power-one oracle inequality that relates a prediction-like measure (left-hand side) with the complexity of the problem (right-hand side).
Since we seek bounds for~$\Delta_{\complementaryparameter}$ and~$\boldsymbol{\Delta}_{\modifiedvariogram}$ directly,
and since the right-hand side still involves the estimator,
the oracle inequality is of limited value by itself but rather a stepping stone in proving our Theorem~\ref{regtheory}.
However, interestingly, the stated oracle inequality is a bound that does not require any assumptions---neither on the data~$\argument_1,\dots,\argument_{\numberofobservations}$ nor on the tuning parameter~\tuningparameter.

\section{Proofs}

To simplify the further treatment, we merge the parameters.
Assume all the pdfs are differentiable in~$\mathbb{R}_{+}^{\nbrdim}$, and also assume that the data pdf~$\empiricaldensityF{\argument}$ is differentiable in~$\mathbb{R}_{+}^{\nbrdim}$, the expectations~$\E_{\argument}\left\{\normB{\scorefunction[\argument ; \complementaryparameter,\modifiedvariogram]}^{2}\norm{\argument}^{2}\right\}$ and~$\E_{\argument}\left\{\normB{\scorefunction[\argument]}^{2}\norm{\argument}^{2}\right\}$ are finite for any~$\complementaryparameter,\modifiedvariogram$, and~$\empiricaldensityF{\argument} \scorefunction[\argument ; \complementaryparameter,\modifiedvariogram] \argumentE_{i}^{2}$ goes to zero for any~$i$ and~$\complementaryparameter,\modifiedvariogram$ when~$\norm{\argument} \rightarrow \infty$ or $\norm{\argument} \rightarrow 0$.
Also define that the vector $\mones\in\R^{\nbrdim}$ consists of $1$'s in each element: $\mones_{i}=1$.

\subsection{Proof of Lemma~\ref{res:margin}}

\begin{proof}[Proof of Lemma~\ref{res:margin}]
	The definition of the objective function $\objectivefunctionI$ yields
	\begin{multline*}
		\objectivefunctionI[\complementaryparameter,\modifiedvariogram,\argument]~=~\objectivefunctionI[\complementaryparameter',\modifiedvariogram',\argument]
		+\inprodB{\nabla\objectivefunctionI[\complementaryparameter',\modifiedvariogram',\argument]}{(\Delta_{\complementaryparameter},\boldsymbol{\Delta}_{\modifiedvariogram})}\\
		+\normtwosB{\bigl(\Delta_{\complementaryparameter}-\boldsymbol{\Delta}_{\modifiedvariogram}\log[\argument]\bigr)\otimes\datafunctionA}
	\end{multline*}
	for every $\argument\in\R^{\nbrdim}$.
	Taking sums over the samples on both sides yields the first claim of the lemma.
	
	Taking instead expectations on both sides and using the linearity of inner products and integrals then gives us
	\begin{multline*}
		\E_{\argument}\bigl[\objectivefunctionI[\complementaryparameter,\modifiedvariogram,\argument]\bigr]~=~\E_{\argument}\bigl[\objectivefunctionI[\complementaryparameter',\modifiedvariogram',\argument]\bigr]
		+\inprodB{\E_{\argument}\bigl[\nabla\objectivefunctionI[\complementaryparameter',\modifiedvariogram',\argument]\bigr]}{(\Delta_{\complementaryparameter},\boldsymbol{\Delta}_{\modifiedvariogram})}\\
		+\E_{\argument}\Bigl[\normtwosB{\bigl(\Delta_{\complementaryparameter}-\boldsymbol{\Delta}_{\modifiedvariogram}\log[\argument]\bigr)\otimes\datafunctionA}\Bigr]\,.
	\end{multline*}
	Since the objective function~$\objectivefunctionI$ is smooth in~$\complementaryparameter$ and $\modifiedvariogram$,
	we can use Leibniz' rule to interchange differentiation and integration,
	providing the second part of the lemma.
\end{proof}

\subsection{Proof of Lemma~\ref{res:cone}}

\begin{proof}[Proof of Lemma~\ref{res:cone}]
	By Definition~\ref{regscore} of the estimator,
	it holds that
	\begin{equation*}
		\sum_{\indexsample=1}^{\numberofobservations}\objectivefunctionI[\complementaryparameterest,\modifiedvariogramLest,\argumenti]+\tuningparameter\priorF{\complementaryparameterest,\modifiedvariogramLest}~\leq~\sum_{\indexsample=1}^{\numberofobservations}\objectivefunctionI[\complementaryparameter,\modifiedvariogramL,\argumenti]+\tuningparameter\priorF{\complementaryparameter,\modifiedvariogramL}\,,
	\end{equation*}
	that is,
	\begin{equation*}
		\tuningparameter\priorF{\complementaryparameter,\modifiedvariogramL} -\tuningparameter\priorF{\complementaryparameterest,\modifiedvariogramLest}~\geq~\sum_{\indexsample=1}^{\numberofobservations}\objectivefunctionI[\complementaryparameterest,\modifiedvariogramLest,\argumenti]-\sum_{\indexsample=1}^{\numberofobservations}\objectivefunctionI[\complementaryparameter,\modifiedvariogramL,\argumenti]\,.
	\end{equation*}
	We then bound either side of this inequality.
	
	First, with Assumption~\ref{ass:sepa},
	\begin{align*}
		&\tuningparameter\priorF{\complementaryparameter,\modifiedvariogramL} -\tuningparameter\priorF{\complementaryparameterest,\modifiedvariogramLest}\\
		&=~\tuningparameter\priorF{\complementaryparameter\sparsityVU,\modifiedvariogramL\sparsityMU}+\tuningparameter\priorF{\complementaryparameter\sparsityVUC,\modifiedvariogramL\sparsityMUC} -\tuningparameter\priorF{\complementaryparameterest\sparsityVU,\modifiedvariogramLest\sparsityMU}-\tuningparameter\priorF{\complementaryparameterest\sparsityVUC,\modifiedvariogramLest\sparsityMUC} \displaybreak[0]\\
		&=~\tuningparameter\priorF{\complementaryparameter\sparsityVU,\modifiedvariogramL\sparsityMU} -\tuningparameter\priorF{\complementaryparameterest\sparsityVU,\modifiedvariogramLest\sparsityMU}-\tuningparameter\priorF{({\Delta}_{\complementaryparameter})\sparsityVUC,(\boldsymbol{\Delta}_{\modifiedvariogram})\sparsityMUC}\\
		&\leq~\tuningparameter\priorF{({\Delta}_{\complementaryparameter})\sparsityVU,(\boldsymbol{\Delta}_{\modifiedvariogram})\sparsityMU}-\tuningparameter\priorF{({\Delta}_{\complementaryparameter})\sparsityVUC,(\boldsymbol{\Delta}_{\modifiedvariogram})\sparsityMUC}\,.
	\end{align*}
	
	On the other hand,
	using Lemma~\ref{res:margin} with $\complementaryparameter'=\complementaryparameterest$ and $\modifiedvariogram'=\modifiedvariogramest$, and using \cite[Lemma~A.1]{Zhuang2018} (recall that we use~$\modifiedvariogramL$ and~$\modifiedvariogram$ interchangeably)
	\begin{align*}
		&\sum_{\indexsample=1}^{\numberofobservations}\objectivefunctionI[\complementaryparameterest,\modifiedvariogramLest,\argumenti]-\sum_{\indexsample=1}^{\numberofobservations}\objectivefunctionI[\complementaryparameter,\modifiedvariogramL,\argumenti]\\
		&=~\inprodBBB{\sum_{\indexsample=1}^{\numberofobservations}\nabla\objectivefunctionI[\complementaryparameter,\modifiedvariogram,\argumenti]}{({\Delta}_{\complementaryparameter},\boldsymbol{\Delta}_{\modifiedvariogram})}
		+\sum_{\indexsample=1}^{\numberofobservations}\Bigl[\normtwosB{\bigl({\Delta}_{\complementaryparameter}-\boldsymbol{\Delta}_{\modifiedvariogram}\log[\argumenti]\bigr)\otimes\datafunction[\argumenti]}\Bigr]\\
		&\geq~\inprodBBB{\sum_{\indexsample=1}^{\numberofobservations}\nabla\objectivefunctionI[\complementaryparameter,\modifiedvariogram,\argumenti]}{({\Delta}_{\complementaryparameter},\boldsymbol{\Delta}_{\modifiedvariogram})}\\
		&\geq~-\priorDFBBBB{\sum_{\indexsample=1}^{\numberofobservations}\nabla\objectivefunctionI[\complementaryparameter,\modifiedvariogram,\argumenti]}\priorF{{\Delta}_{\complementaryparameter},\boldsymbol{\Delta}_{\modifiedvariogram}}\\
		&=~-\tuningparameterOA\priorF{{\Delta}_{\complementaryparameter},\boldsymbol{\Delta}_{\modifiedvariogram}}\\
		&\geq~-\frac{\tuningparameter}{2}\priorF{{\Delta}_{\complementaryparameter},\boldsymbol{\Delta}_{\modifiedvariogram}}\\
		&=~-\frac{\tuningparameter}{2}\priorF{({\Delta}_{\complementaryparameter})\sparsityVU,(\boldsymbol{\Delta}_{\modifiedvariogram})\sparsityMU}-\frac{\tuningparameter}{2}\priorF{({\Delta}_{\complementaryparameter})\sparsityVUC,(\boldsymbol{\Delta}_{\modifiedvariogram})\sparsityMUC}\,.
	\end{align*}
	
	Collecting the pieces yields the desired result.
\end{proof}

\subsection{Proof of Lemma~\ref{res:oe}}

\begin{proof}[Proof of Lemma~\ref{res:oe}]
	The proof essentially combines Lemma~\ref{res:margin} with the definition of the estimator.
	
	The first part of Lemma~\ref{res:margin} with $\complementaryparameter'=\complementaryparameterest$ and $\modifiedvariogram'=\modifiedvariogramest$ yields
	\begin{multline*}
		\sum_{\indexsample=1}^{\numberofobservations}\objectivefunctionI[\complementaryparameterest,\modifiedvariogramest,\argumenti]~=~\sum_{\indexsample=1}^{\numberofobservations}\objectivefunctionI[\complementaryparameter,\modifiedvariogram,\argumenti]
		+\inprodBBB{\sum_{\indexsample=1}^{\numberofobservations}\nabla\objectivefunctionI[\complementaryparameter,\modifiedvariogram,\argumenti]}{(\Delta_{\complementaryparameter},\boldsymbol{\Delta}_{\modifiedvariogram})}\\
		+\sum_{\indexsample=1}^{\numberofobservations}\normtwosB{\bigl(\Delta_{\complementaryparameter}-\boldsymbol{\Delta}_{\modifiedvariogram}\log[\argumenti]\bigr)\otimes\datafunctionAi}\,,
	\end{multline*}
	which is equivalent to (add $\tuningparameter\priorF{\complementaryparameterest,\modifiedvariogramLest}$ to both sides of the equality)
	\begin{multline*}
		\sum_{\indexsample=1}^{\numberofobservations}\objectivefunctionI[\complementaryparameterest,\modifiedvariogramest,\argumenti]+\tuningparameter\priorF{\complementaryparameterest,\modifiedvariogramLest}~=~\sum_{\indexsample=1}^{\numberofobservations}\objectivefunctionI[\complementaryparameter,\modifiedvariogram,\argumenti]
		+\tuningparameter\priorF{\complementaryparameterest,\modifiedvariogramLest}+\inprodBBB{\sum_{\indexsample=1}^{\numberofobservations}\nabla\objectivefunctionI[\complementaryparameter,\modifiedvariogram,\argumenti]}{(\Delta_{\complementaryparameter},\boldsymbol{\Delta}_{\modifiedvariogram})}\\
		+\sum_{\indexsample=1}^{\numberofobservations}\normtwosB{\bigl(\Delta_{\complementaryparameter}-\boldsymbol{\Delta}_{\modifiedvariogram}\log[\argumenti]\bigr)\otimes\datafunctionAi}\,.
	\end{multline*}
	Using then Definition~\ref{regscore} of the estimator gives
	\begin{multline*}
		\sum_{\indexsample=1}^{\numberofobservations}\objectivefunctionI[\complementaryparameter,\modifiedvariogram,\argumenti]+\tuningparameter\priorF{\complementaryparameter,\modifiedvariogramL}~\geq~\sum_{\indexsample=1}^{\numberofobservations}\objectivefunctionI[\complementaryparameter,\modifiedvariogram,\argumenti]
		+\tuningparameter\priorF{\complementaryparameterest,\modifiedvariogramLest}+\inprodBBB{\sum_{\indexsample=1}^{\numberofobservations}\nabla\objectivefunctionI[\complementaryparameter,\modifiedvariogram,\argumenti]}{(\Delta_{\complementaryparameter},\boldsymbol{\Delta}_{\modifiedvariogram})}\\
		+\sum_{\indexsample=1}^{\numberofobservations}\normtwosB{\bigl(\Delta_{\complementaryparameter}-\boldsymbol{\Delta}_{\modifiedvariogram}\log[\argumenti]\bigr)\otimes\datafunctionAi}\,,
	\end{multline*}
	which can be simplified to
	\begin{multline*}
		\tuningparameter\priorF{\complementaryparameter,\modifiedvariogramL}-\tuningparameter\priorF{\complementaryparameterest,\modifiedvariogramLest}~\geq~\inprodBBB{\sum_{\indexsample=1}^{\numberofobservations}\nabla\objectivefunctionI[\complementaryparameter,\modifiedvariogram,\argumenti]}{(\Delta_{\complementaryparameter},\boldsymbol{\Delta}_{\modifiedvariogram})}\\
		+\sum_{\indexsample=1}^{\numberofobservations}\normtwosB{\bigl(\Delta_{\complementaryparameter}-\boldsymbol{\Delta}_{\modifiedvariogram}\log[\argumenti]\bigr)\otimes\datafunctionAi}\,.
	\end{multline*}
	Next, using H\"older's inequality \cite[Lemma~A.1]{Zhuang2018} on the right-hand side, we find
	\begin{multline*}
		\tuningparameter\priorF{\complementaryparameter,\modifiedvariogramL}-\tuningparameter\priorF{\complementaryparameterest,\modifiedvariogramLest}~\geq~-\priorDFBBBB{\sum_{\indexsample=1}^{\numberofobservations}\nabla\objectivefunctionI[\complementaryparameter,\modifiedvariogram,\argumenti]}\priorF{\Delta_{\complementaryparameter},\boldsymbol{\Delta}_{\modifiedvariogram}}\\
		+\sum_{\indexsample=1}^{\numberofobservations}\normtwosB{\bigl(\Delta_{\complementaryparameter}-\boldsymbol{\Delta}_{\modifiedvariogram}\log[\argumenti]\bigr)\otimes\datafunctionAi}\,,
	\end{multline*}
	that is,
	\begin{equation*}
		\tuningparameter\priorF{\complementaryparameter,\modifiedvariogramL}-\tuningparameter\priorF{\complementaryparameterest,\modifiedvariogramLest}~\geq~-\tuningparameterOA\priorF{\Delta_{\complementaryparameter},\boldsymbol{\Delta}_{\modifiedvariogram}}
		+\sum_{\indexsample=1}^{\numberofobservations}\normtwosB{\bigl(\Delta_{\complementaryparameter}-\boldsymbol{\Delta}_{\modifiedvariogram}\log[\argumenti]\bigr)\otimes\datafunctionAi}\,.
	\end{equation*}
	Using the triangle inequality on the left-hand side gives
	\begin{equation*}
		\tuningparameter\priorF{\Delta_{\complementaryparameter},\boldsymbol{\Delta}_{\modifiedvariogram}}~\geq~-\tuningparameterOA\priorF{\Delta_{\complementaryparameter},\boldsymbol{\Delta}_{\modifiedvariogram}}
		+\sum_{\indexsample=1}^{\numberofobservations}\normtwosB{\bigl(\Delta_{\complementaryparameter}-\boldsymbol{\Delta}_{\modifiedvariogram}\log[\argumenti]\bigr)\otimes\datafunctionAi}\,,
	\end{equation*}
	which can be written in the desired form:
	\begin{equation*}
		\sum_{\indexsample=1}^{\numberofobservations}\normtwosB{\bigl(\Delta_{\complementaryparameter}-\boldsymbol{\Delta}_{\modifiedvariogram}\log[\argumenti]\bigr)\otimes\datafunctionAi}~\leq~\bigl(\tuningparameter+\tuningparameterOA\bigr)\priorF{\Delta_{\complementaryparameter},\boldsymbol{\Delta}_{\modifiedvariogram}}\,.
	\end{equation*}
\end{proof}

\subsection{Proof of Lemma~\ref{reparametrization}}

\begin{proof}[Proof of Lemma~\ref{reparametrization}]
	The proof is a tedious but straightforward calculation.
	
	\emph{Step 1:}
	We first show that 
	\begin{multline*}
		\densitytradF{\argument;\variogram}~=~\frac{c'_{\variogram}}{\partitiontrad} \frac{1}{\argumenthigh}\Biggl(\prod_{\indexelement=1}^{\nbrdim}\frac{1}{\argumentE_{\indexelement}}\Biggr) \exp\Bigl[ -\frac{1}{2} \log[\argumenthighM/\argumenthigh]\tp\covariance ^{-1} \log[\argumenthighM/\argumenthigh]  \Bigr] \\
		\times\exp\Bigl[ - \frac{1}{2}(\variogram_{-\indexhigh,\indexhigh})\tp  \covariance ^{-1}\log[\argumenthighM/\argumenthigh] \Bigr]
	\end{multline*}
	with
	\begin{equation*}
		c'_{\variogram}~\deq~\frac{1}{\sqrt{(2\pi)^{\nbrdim-1} \det[\covariance]}}   \exp\Bigl[-\frac{1}{8}(\variogram_{-\indexhigh,\indexhigh})\tp  \covariance ^{-1} \variogram_{-\indexhigh,\indexhigh} \Bigr]\,.
	\end{equation*}

	Recall from Equation~\eqref{eq:spectral-density} that
	\begin{equation*}
		\densitytradF{\argument;\variogram}~=~\frac{1}{\partitiontrad} \frac{1}{\argumenthigh}\Biggl(\prod_{\indexelement=1}^{\nbrdim}\frac{1}{\argumentE_{\indexelement}}\Biggr)\normaldensityFB{\log[\argumenthighM/\argumenthigh];-\variogram_{-\indexhigh,\indexhigh}/2,\covariance}\,.
	\end{equation*}
	Using the definition of the multivariate Gaussian distribution then gives
	\begin{multline*}
		\densitytradF{\argument;\variogram}~=~\frac{1}{\partitiontrad} \frac{1}{\argumenthigh}\Biggl(\prod_{\indexelement=1}^{\nbrdim}\frac{1}{\argumentE_{\indexelement}}\Biggr) \frac{1}{\sqrt{(2\pi)^{\nbrdim-1} \det[\covariance]}}\\ 
		\times\exp\Bigl[-\frac{1}{2}(\log[\argumenthighM/\argumenthigh]
		+\variogram_{-\indexhigh,\indexhigh}/2)\tp \covariance ^{-1}(\log[\argumenthighM/\argumenthigh]
		+\variogram_{-\indexhigh,\indexhigh}/2)\Bigr]\,.
	\end{multline*}
	Next, expanding the terms in the exponent yields
	\begin{multline*}
		\densitytradF{\argument;\variogram}~=~\frac{1}{\partitiontrad} \frac{1}{\argumenthigh}\Biggl(\prod_{\indexelement=1}^{\nbrdim}\frac{1}{\argumentE_{\indexelement}}\Biggr)\frac{1}{\sqrt{(2\pi)^{\nbrdim-1} \det[\covariance]}} \exp\Bigl[-\frac{1}{2}\log[\argumenthighM/\argumenthigh]\tp \covariance ^{-1} \log[\argumenthighM/\argumenthigh] \Bigr] \\
		\times\exp\Bigl[ -\frac{1}{2} \log[\argumenthighM/\argumenthigh]\tp \covariance ^{-1} \variogram_{-\indexhigh,\indexhigh}/2 -\frac{1}{2} (\variogram_{-\indexhigh,\indexhigh}/2)\tp \covariance ^{-1}\log[\argumenthighM/\argumenthigh] \Bigr] \\
		\times \exp\Bigl[-\frac{1}{2}(\variogram_{-\indexhigh,\indexhigh}/2)\tp \covariance ^{-1} \variogram_{-\indexhigh,\indexhigh}/2 \Bigr]\,.
	\end{multline*}
	Basic linear algebra and the fact that $\covariance$ is symmetric then yields
	\begin{multline*}
		\densitytradF{\argument;\variogram}~=~\frac{1}{\partitiontrad} \frac{1}{\argumenthigh}\Biggl(\prod_{\indexelement=1}^{\nbrdim}\frac{1}{\argumentE_{\indexelement}}\Biggr)\frac{1}{\sqrt{(2\pi)^{\nbrdim-1} \det[\covariance]}} \exp\Bigl[-\frac{1}{2}\log[\argumenthighM/\argumenthigh]\tp \covariance ^{-1} \log[\argumenthighM/\argumenthigh] \Bigr] \\
		\times\exp\Bigl[  -\frac{1}{2}(\variogram_{-\indexhigh,\indexhigh})\tp \covariance ^{-1}\log[\argumenthighM/\argumenthigh]   \Bigr] \exp\Bigl[-\frac{1}{8}(\variogram_{-\indexhigh,\indexhigh})\tp  \covariance ^{-1} \variogram_{-\indexhigh,\indexhigh} \Bigr]\,.
	\end{multline*}
	Hence, using the definition of $c'_{\variogram}$,
	\begin{multline*}
		\densitytradF{\argument;\variogram}~=~\frac{c'_{\variogram}}{\partitiontrad} \frac{1}{\argumenthigh}\Biggl(\prod_{\indexelement=1}^{\nbrdim}\frac{1}{\argumentE_{\indexelement}}\Biggr)\exp\Bigl[-\frac{1}{2}\log[\argumenthighM/\argumenthigh]\tp \covariance ^{-1} \log[\argumenthighM/\argumenthigh] \Bigr] \\
		\times\exp\Bigl[  -\frac{1}{2}(\variogram_{-\indexhigh,\indexhigh})\tp \covariance ^{-1}\log[\argumenthighM/\argumenthigh]   \Bigr]\,,
	\end{multline*}
	as desired.

	\emph{Step 2:}
	We then show that 
	\begin{equation*}
		\densitytradF{\argument;\variogram}~=~\frac{c'_{\variogram}}{\partitiontrad} \Biggl(\prod_{\indexelement=1}^{\nbrdim}\frac{1}{\argumentE_{\indexelement}}\Biggr) \exp\Bigl[\complementaryparameter\tp\log[\argument] -\frac{1}{2} \log[\argumenthighM/\argumenthigh]\tp \covariance ^{-1} \log[\argumenthighM/\argumenthigh]  \Bigr] \,.
	\end{equation*}
	
	We essentially need to rewrite the second exponent of the result of Step~1.
	We find with the definition of~\complementaryparameter\ that
	\begin{align*}
		&- \frac{1}{2}(\variogram_{-\indexhigh,\indexhigh})\tp  \covariance ^{-1}\log[\argumenthighM/\argumenthigh]-\log[\argumenthigh]\\
		&=~\sum_{\indexelementP\neq\indexhigh}\Bigl(- \frac{1}{2}\covariance ^{-1}\variogram_{-\indexhigh,\indexhigh}\Bigr)_{\indexelementP}\log[\argumentE_{\indexelementP}/\argumenthigh]-\log[\argumenthigh]\\
		&=~\sum_{\indexelementP\neq\indexhigh}\Bigl(- \frac{1}{2}\covariance\inv\variogram_{-\indexhigh,\indexhigh}\Bigr)_{\indexelementP}\bigl(\log[\argumentE_{\indexelementP}]-\log[\argumenthigh]\bigr)-\log[\argumenthigh]\\
		&=~\sum_{\indexelementP\neq\indexhigh}\Bigl(- \frac{1}{2}\covariance\inv\variogram_{-\indexhigh,\indexhigh}\Bigr)_{\indexelementP}\log[\argumentE_{\indexelementP}]+\biggl(\sum_{\indexelementP\neq\indexhigh}\Bigl(\frac{1}{2}\covariance\inv\variogram_{-\indexhigh,\indexhigh}\Bigr)_{\indexelementP}\biggr)\log[\argumenthigh]-\log[\argumenthigh]\\
		&=~\sum_{\indexelementP\neq\indexhigh}\complementaryparameterE_{\indexelementP}\log[\argumentE_{\indexelementP}]+\complementaryparameterE_{\indexhigh}\log[\argumenthigh]\\
		&=~\complementaryparameter\tp\log[\argument]\,.
	\end{align*}
	Combining this result with Step~1 yields the desired statement.
	
	\emph{Step 3:}
	We finally show that
	\begin{equation*}
		\densitytradF{\argument;\variogram}~=~\frac{c'_{\variogram}}{\partitiontrad} \Biggl(\prod_{\indexelement=1}^{\nbrdim}\frac{1}{\argumentE_{\indexelement}}\Biggr) \exp\Bigl[\complementaryparameter\tp\log[\argument]-\frac{1}{2}\log[\argument]\tp \modifiedvariogram\log[\argument] \Bigr] \,.
	\end{equation*}
	
	We need to rewrite the second exponent of the result of Step~2.
	Basic algebra and the definition of $\modifiedvariogram$ gives us
	\begin{align*}
		&-\frac{1}{2} \log[\argumenthighM/\argumenthigh]\tp \covariance ^{-1} \log[\argumenthighM/\argumenthigh]\\
		&=~-\frac{1}{2}\sum_{\indexelementP,\indexelementPP=1}^{\nbrdim-1}\log[(\argumenthighM)_{\indexelementP}/\argumenthigh] (\covarianceE\inv)_{\indexelementP\indexelementPP}\log[(\argumenthighM)_{\indexelementPP}/\argumenthigh] \displaybreak[0]\\
		&=~-\frac{1}{2}\sum_{\indexelementP,\indexelementPP=1}^{\nbrdim-1}\log[(\argumenthighM)_{\indexelementP}] (\covarianceE\inv)_{\indexelementP\indexelementPP}\log[(\argumenthighM)_{\indexelementPP}]+\frac{1}{2}\sum_{\indexelementP,\indexelementPP=1}^{\nbrdim-1}\log[(\argumenthighM)_{\indexelementP}](\covarianceE\inv)_{\indexelementP\indexelementPP}\log[\argumenthigh]\\
		&~~~~~+\frac{1}{2}\sum_{\indexelementP,\indexelementPP=1}^{\nbrdim-1}\log[\argumentE_{\indexhigh}](\covarianceE\inv)_{\indexelementP\indexelementPP}\log[(\argumenthighM)_{\indexelementPP}]-\frac{1}{2}\sum_{\indexelementP,\indexelementPP=1}^{\nbrdim-1}\log[\argumentE_{\indexhigh}](\covarianceE\inv)_{\indexelementP\indexelementPP}\log[\argumenthigh] \displaybreak[0]\\
		&=~-\frac{1}{2}\sum_{\indexelementP,\indexelementPP=1}^{\nbrdim-1}\log[(\argumenthighM)_{\indexelementP}] (\covarianceE\inv)_{\indexelementP\indexelementPP}\log[(\argumenthighM)_{\indexelementPP}]-\frac{1}{2}\sum_{\indexelementP=1}^{\nbrdim-1}\log[(\argumenthighM)_{\indexelementP}]\Bigl(-\sum_{\indexelementPP=1}^{\nbrdim-1}(\covarianceE\inv)_{\indexelementP\indexelementPP}\Bigr)\log[\argumenthigh]\\
		&~~~~~-\frac{1}{2}\sum_{\indexelementPP=1}^{\nbrdim-1}\log[\argumentE_{\indexhigh}]\Bigl(-\sum_{\indexelementP=1}^{\nbrdim-1}(\covarianceE\inv)_{\indexelementP\indexelementPP}\Bigr)\log[(\argumenthighM)_{\indexelementPP}]-\frac{1}{2}\log[\argumentE_{\indexhigh}]\Bigl(\sum_{\indexelementP,\indexelementPP=1}^{\nbrdim-1}(\covarianceE\inv)_{\indexelementP\indexelementPP}\Bigr)\log[\argumenthigh]\\
		&=~-\frac{1}{2}\sum_{\substack{\indexelementP,\indexelementPP\in\{1,\dots,\nbrdim\}\\\indexelementP,\indexelementPP\neq\indexhigh}}\log[\argumentE_{\indexelementP}]\log[\argumentE_{\indexelementPP}]\times 
		\begin{cases}
			(\covarianceE\inv)_{\indexelementP\indexelementPP}~~~&\text{for}~\indexelementP,\indexelementPP<\indexhigh\\
			(\covarianceE\inv)_{(\indexelementP-1)\indexelementPP}~~~&\text{for}~\indexelementP>\indexelement;\indexelementPP<\indexhigh\\
			(\covarianceE\inv)_{\indexelementP(\indexelementPP-1)}~~~&\text{for}~\indexelementP<\indexelement;\indexelementPP>\indexhigh\\
			(\covarianceE\inv)_{(\indexelementP-1)(\indexelementPP-1)}~~~&\text{for}~\indexelementP,\indexelementPP>\indexhigh
		\end{cases}\\
		&~~~~~-\frac{1}{2}\sum_{\substack{\indexelementP\in\{1,\dots,\nbrdim\}\\\indexelementP\neq\indexhigh}}\log[\argumentE_{\indexelementP}]\log[\argumenthigh]\times
		\begin{cases}
			-\sum_{\indexelementPP=1}^{\nbrdim-1}(\covarianceE\inv)_{\indexelementP\indexelementPP}~~~&\text{for}~\indexelementP<\indexhigh\\
			-\sum_{\indexelementPP=1}^{\nbrdim-1}(\covarianceE\inv)_{(\indexelementP+1)\indexelementPP}~~~&\text{for}~\indexelementP>\indexhigh
		\end{cases}\\
		&~~~~~-\frac{1}{2}\sum_{\substack{\indexelementPP\in\{1,\dots,\nbrdim\}\\\indexelementPP\neq\indexhigh}}\log[\argumentE_{\indexhigh}]\log[\argumentE_{\indexelementPP}] \times
		\begin{cases}
			-\sum_{\indexelementP=1}^{\nbrdim-1}(\covarianceE\inv)_{\indexelementP\indexelementPP}~~~&\text{for}~\indexelementPP<\indexhigh\\
			-\sum_{\indexelementP=1}^{\nbrdim-1}(\covarianceE\inv)_{\indexelementP(\indexelementPP-1)}~~~&\text{for}~\indexelementPP>\indexhigh
		\end{cases}\\
		&~~~~~-\frac{1}{2}\log[\argumentE_{\indexhigh}]\Bigl(\sum_{\indexelementP,\indexelementPP=1}^{\nbrdim-1}(\covarianceE\inv)_{\indexelementP\indexelementPP}\Bigr)\log[\argumenthigh]\\
		&=~-\frac{1}{2}\log[\argument]\tp \modifiedvariogram \log[\argument]
	\end{align*}
	Combining this result with Step~2 yields the claimed statements.
\end{proof}

\subsection{Proof of Lemma~\ref{scorefunctions}}
\begin{proof}[Proof of Lemma~\ref{scorefunctions}]
	We can easily derive
	\begin{align*}
		\scorefunctionE_{\indexelement}[\argument;\complementaryparameter,\modifiedvariogramL]
		~&=~ \frac{\partial \log\bigl[\densityF{\argument;\complementaryparameter,\modifiedvariogramL}\bigr]}{\partial \argumentE_{\indexelement}}\\
		&=- \frac{1}{\argumentE_{\indexelement}} +\frac{\complementaryparameterE_{\indexelement}}{\argumentE_{\indexelement}}-\frac{1}{2}\frac{\partial}{\partial \argumentE_{\indexelement}}\log[\argument]\tp (\modifiedvariogramL+\modifiedvariogramL\tp\varioext)\log[\argument] \displaybreak[0]\\
		&=- \frac{1}{\argumentE_{\indexelement}} +\frac{\complementaryparameterE_{\indexelement}}{\argumentE_{\indexelement}}-\frac{\bigl((\modifiedvariogramL+\modifiedvariogramL\tp\varioext)\log[\argument]\bigr)_{\indexelement}}{\argumentE_{\indexelement}}\\
		&=~ \frac{\complementaryparameterE_{\indexelement}-1-\bigl((\modifiedvariogramL+\modifiedvariogramL\tp\varioext)\log[\argument]\bigr)_{\indexelement}}{\argumentE_{\indexelement}}\,,
	\end{align*}
	as desired.
\end{proof}

\subsection{Proof of Theorem~\ref{scorematchinggeneral}}
\begin{proof}[Proof of Theorem~\ref{scorematchinggeneral}]
	
	By the definition of the objective function in~\eqref{scalingscorematching} and the linearity of integrals, we get
	\begin{align*}
		\objectivefunction [\complementaryparameter,\modifiedvariogramL]
		~&=~\frac{1}{2} \int_{\maxargumentS} \empiricaldensityF{\argument}\normtwoB{\scorefunction [\argument] \otimes\argument\otimes\weightA-\scorefunction[\argument ; \complementaryparameter,\modifiedvariogramL] \otimes\argument\otimes\weightA}^{2} \mathrm{~d} \argument\\
		&=~\frac{1}{2} \int_{\maxargumentS} \empiricaldensityF{\argument}\normtwosB{\scorefunction[\argument ; \complementaryparameter,\modifiedvariogramL] \otimes\argument\otimes\weightA} \mathrm{~d} \argument\\
		&~~~~~-\int_{\maxargumentS} \empiricaldensityF{\argument}\inprod{\scorefunction [\argument] \otimes \argument \otimes\weightA \otimes \scorefunction[\argument ; \complementaryparameter,\modifiedvariogramL]\otimes\argument\otimes\weightA}{\mones}\mathrm{~d} \argument+\text{const.}\,,
	\end{align*}
	where ``$\text{const}.$'' is independent of the parameters.
	
	We now treat the last two terms in order.
	For the first term, we find
	\begin{align*}
		&\frac{1}{2} \int_{\maxargumentS} \empiricaldensityF{\argument}\normtwosB{\scorefunction[\argument ; \complementaryparameter,\modifiedvariogramL] \otimes\argument\otimes\weightA} \mathrm{~d} \argument\\
		&=~\frac{1}{2} \int_{\maxargumentS} \empiricaldensityF{\argument}\sum_{\indexelement=1}^{\nbrdim}\underbrace{(\weightEF{\argumentE_{\indexelement}})^2}_{(\datafunctionE[\argumentE_{\indexelement}])^2}\Bigl(\complementaryparameterE_{\indexelement}-1-\bigl((\modifiedvariogramL+\modifiedvariogramL\tp\varioext)\log[\argument]\bigr)_{\indexelement}\Bigr)^2 \mathrm{~d} \argument\tj{by Lemma~\ref{scorefunctions}} \displaybreak[0]\\
		&=~\frac{1}{2} \int_{\maxargumentS} \empiricaldensityF{\argument}\normtwosB{\bigl(\complementaryparameter-\mones-(\modifiedvariogramL+\modifiedvariogramL\tp\varioext)\log[\argument]\bigr)\otimes\datafunction[\argument]} \mathrm{~d} \argument\,.
	\end{align*}
	
	For the second term, we find
	\begin{align*}
		&-\int_{\maxargumentS} \empiricaldensityF{\argument}\inprod{\scorefunction [\argument] \otimes\argument\otimes\weightA \otimes \scorefunction[\argument ; \complementaryparameter,\modifiedvariogramL]\otimes\argument\otimes\weightA}{\mones}\mathrm{~d} \argument\\
		&=~-\int_{\maxargumentS} \empiricaldensityF{\argument}\inprodB{(\nabla_{\argument} \log \empiricaldensityF{\argument}) \otimes \argument\otimes\weightA \otimes \scorefunction[\argument ; \complementaryparameter,\modifiedvariogramL]\otimes\argument\otimes\weightA}{\mones}\mathrm{~d} \argument\\
		&=~-\int_{\maxargumentS} \inprodB{(\nabla_{\argument}\empiricaldensityF{\argument})\otimes \argument\otimes\weightA \otimes \scorefunction[\argument ; \complementaryparameter,\modifiedvariogramL]\otimes \argument\otimes\weightA}{\mones}\mathrm{~d} \argument\\
		&=~\int_{\maxargumentS} \inprodB{\empiricaldensityF{\argument}\cdot\nabla_{\argument}\otimes\bigl(\argument\otimes\weightA \otimes \scorefunction[\argument ; \complementaryparameter,\modifiedvariogramL]\otimes\argument\otimes\weightA\bigr)}{\mones}\mathrm{~d} \argument\tj{integration by parts---cf.~Assumption~\ref{ass:dgd}}\\
		&=~\int_{\maxargumentS} \empiricaldensityF{\argument}\sum_{\indexelement=1}^{\nbrdim}\frac{\partial}{\partial \argumentE_{\indexelement}}\biggl(\argumentE_{\indexelement}(\weightEF{\argumentE_{\indexelement}})^2 \Bigl(\complementaryparameterE_{\indexelement}-1-\bigl((\modifiedvariogramL+\modifiedvariogramL\tp\varioext)\log[\argument]\bigr)_{\indexelement}\Bigr)\biggr)\mathrm{~d} \argument\\
		&=~\int_{\maxargumentS} \empiricaldensityF{\argument}\sum_{\indexelement=1}^{\nbrdim}\biggl( \underbrace{\bigl((\weightEF{\argumentE_{\indexelement}})^2+2\argumentE_{\indexelement}\weightE'[\argumentE_{\indexelement}]\weightEF{\argumentE_{\indexelement}}\bigr)}_{\frac{1}{2}  \datafunctionPE[\argumentE_{\indexelement}]} \Bigl(\complementaryparameterE_{\indexelement}-1-\bigl((\modifiedvariogramL+\modifiedvariogramL\tp\varioext)\log[\argument]\bigr)_{\indexelement}\Bigr)\biggr)\mathrm{~d} \argument\\
		&~~~~~-\int_{\maxargumentS} \empiricaldensityF{\argument}\sum_{\indexelement=1}^{\nbrdim}\biggl((\weightEF{\argumentE_{\indexelement}})^2 \varioextE_{\indexelement\indexelement}\biggr)\mathrm{~d} \argument \displaybreak[0]\\
		&=~\frac{1}{2}\int_{\maxargumentS} \empiricaldensityF{\argument}\bigl(\complementaryparameter-\mones-(\modifiedvariogramL+\modifiedvariogramL\tp\varioext)\log[\argument]\bigr)\tp\datafunctionP[\argument]\mathrm{~d} \argument\\
		&~~~~~-\int_{\maxargumentS} \empiricaldensityF{\argument}\sum_{\indexelement,\indexelementP=1}^{\nbrdim}\biggl(\underbrace{\diag\bigl[(\weightEF{\argumentE_{\indexelement}})^2\bigr]_{\indexelement\indexelementP}}_{\frac{1}{2} (\datamatrix[\argument])_{\indexelement\indexelementP}}\bigl(\modifiedvariogramL+\modifiedvariogramL\tp \varioext\bigr)_{\indexelementP\indexelement}\biggr)\mathrm{~d} \argument \displaybreak[0]\\
		&=~\frac{1}{2}\int_{\maxargumentS} \empiricaldensityF{\argument}\bigl(\complementaryparameter-\mones-(\modifiedvariogramL+\modifiedvariogramL\tp\varioext)\log[\argument]\bigr)\tp\datafunctionP[\argument]\mathrm{~d} \argument\\
		&~~~~~- \frac{1}{2} \int_{\maxargumentS} \empiricaldensityF{\argument}\tr\Bigl[\bigl(\modifiedvariogramL+\modifiedvariogramL\tp \varioext\bigr)\datamatrixA\Bigr]\mathrm{~d} \argument\,,
	\end{align*}
	where we have used that
	\begin{align*}
		&\frac{\partial}{\partial \argumentE_{\indexelement}}\bigl((\modifiedvariogramL+\modifiedvariogramL\tp\varioext)\log[\argument]\bigr)_{\indexelement}\\
		&=~(\modifiedvariogramL+\modifiedvariogramL\tp\varioext)_{\indexelement\indexelement}\frac{\partial}{\partial \argumentE_{\indexelement}}\log[\argumentE_{\indexelement}]\\
		&=-~\frac{\varioextE_{\indexelement\indexelement}}{\argumentE_{\indexelement}}\,.
	\end{align*}
\end{proof}

\subsection{Proof of Theorem~\ref{unregtheory}}
\begin{proof}[Proof of Theorem~\ref{unregtheory}]
	Since our objective function is convex, 
	the theorem follows in the case of \hre-distributed data readily from \citet[Theorem~7.77]{Liesebook}.
	In the case of data from the domain of attraction of the \hr model,
	the results can be extended along the lines of \citet{1993deHaan}.
\end{proof}

\subsection{Proof of Theorem~\ref{regtheory}}
\begin{proof}[Proof of Theorem~\ref{regtheory}]
	The proof combines the results of Section~\ref{sec:further}.
	
	Lemma~\ref{res:oe} yields
	\begin{equation*}
		\sum_{\indexsample=1}^{\nbrsamples}\normtwosB{\bigl(\Delta_{\complementaryparameter}-\boldsymbol{\Delta}_{\modifiedvariogram}\log[\argumenti]\bigr)\otimes\datafunctionAi}~\leq~\bigl(\tuningparameter+\tuningparameterOA\bigr)\priorF{\Delta_{\complementaryparameter},\boldsymbol{\Delta}_{\modifiedvariogram}}\,.
	\end{equation*}
	Invoking Assumption~\ref{ass:re},
	this becomes
	\begin{equation*}
		\restrictedeigenvalue\,\nbrsamples\normtwos{(\Delta_{\complementaryparameter})\sparsityVU}+\restrictedeigenvalue\,\nbrsamples\normfs{(\boldsymbol{\Delta}_{\modifiedvariogram})\sparsityMU}~\leq~\bigl(\tuningparameter+\tuningparameterOA\bigr)\priorF{\Delta_{\complementaryparameter},\boldsymbol{\Delta}_{\modifiedvariogram}}\,.
	\end{equation*}
	We then bound the right-hand side as follows:
	\begin{align*}
		&\bigl(\tuningparameter+\tuningparameterOA\bigr)\priorF{\Delta_{\complementaryparameter},\boldsymbol{\Delta}_{\modifiedvariogram}}\\
		&\leq~\frac{3\tuningparameter}{2}\priorF{\Delta_{\complementaryparameter},\boldsymbol{\Delta}_{\modifiedvariogram}}\tj{Assumption~\ref{ass:tp}}\\
		&\leq~\frac{12\tuningparameter}{2}\priorF{(\Delta_{\complementaryparameter})\sparsityVU,(\boldsymbol{\Delta}_{\modifiedvariogram})\sparsityMU}\tj{Lemma~\ref{res:cone}}\\
		&\leq~\frac{12\constprior\sqrt{\abs{\sparsityV}}\tuningparameter}{2}\normtwo{({\Delta}_{\complementaryparameter})\sparsityVU}+\frac{12\constprior\sqrt{\abs{\sparsityM}}\tuningparameter}{2}\normf{(\boldsymbol{\Delta}_{\modifiedvariogram})\sparsityMU}\tj{Assumption~\ref{ass:sepa}}\\
		&\leq~\frac{124\constpriors\abs{\sparsityV}\tuningparameter^2}{4\restrictedeigenvalue\nbrsamples}+\frac{\restrictedeigenvalue\nbrsamples}{4}\normtwo{({\Delta}_{\complementaryparameter})\sparsityVU}^2+\frac{124\constpriors\abs{\sparsityM}\tuningparameter^2}{4\restrictedeigenvalue\nbrsamples}+\frac{\restrictedeigenvalue\nbrsamples}{4}\normfs{(\boldsymbol{\Delta}_{\modifiedvariogram})\sparsityMU}\tj{\citet[Lemma~B.1.3]{Lederer2022}}\\
		&=~\frac{124\constpriors\bigl(\abs{\sparsityV}+\abs{\sparsityM}\bigr)\tuningparameter^2}{4\restrictedeigenvalue\nbrsamples}+\frac{\restrictedeigenvalue\nbrsamples}{4}\normtwo{({\Delta}_{\complementaryparameter})\sparsityVU}^2+\frac{\restrictedeigenvalue\nbrsamples}{4}\normfs{(\boldsymbol{\Delta}_{\modifiedvariogram})\sparsityMU}\,.
	\end{align*}
	
	Combining the two displays yields
	\begin{multline*}
		\restrictedeigenvalue\,\nbrsamples\normtwos{(\Delta_{\complementaryparameter})\sparsityVU}+\restrictedeigenvalue\,\nbrsamples\normfs{(\boldsymbol{\Delta}_{\modifiedvariogram})\sparsityMU}~\\
		\leq~\frac{124\constpriors\bigl(\abs{\sparsityV}+\abs{\sparsityM}\bigr)\tuningparameter^2}{4\restrictedeigenvalue\nbrsamples}+\frac{\restrictedeigenvalue\nbrsamples}{4}\normtwo{({\Delta}_{\complementaryparameter})\sparsityVU}^2+\frac{\restrictedeigenvalue\nbrsamples}{4}\normfs{(\boldsymbol{\Delta}_{\modifiedvariogram})\sparsityMU}\,,
	\end{multline*}
	and, therefore, 
	\begin{equation*}
		\normtwos{({\Delta}_{\complementaryparameter})\sparsityVU}+\normfs{(\boldsymbol{\Delta}_{\modifiedvariogram})\sparsityMU}~\leq~\frac{124\constpriors\bigl(\abs{\sparsityV}+\abs{\sparsityM}\bigr)\tuningparameter^2}{3\restrictedeigenvalues\nbrsamples^2}\,.
	\end{equation*}
	Invoking Lemma~\ref{res:cone} once more yields
	\begin{equation*}
		\normtwos{{\Delta}_{\complementaryparameter}}+\normfs{\boldsymbol{\Delta}_{\modifiedvariogram}}~\leq~\frac{4\cdot124\constpriors\bigl(\abs{\sparsityV}+\abs{\sparsityM}\bigr)\tuningparameter^2}{3\restrictedeigenvalues\nbrsamples^2}\,,
	\end{equation*}
	as desired.
	(We have not made any attempt to optimize the numerical constant.)
\end{proof}

\end{document}